\documentclass[prb,superscriptaddress,twocolumn,amsmath,amssymb]{revtex4-2}
\usepackage{graphicx}
\usepackage{amsmath}
\usepackage{amssymb}
\usepackage{float}
\usepackage{color}

\def\oJ{{\mbox{\boldmath{${\cal J}$}}}}
\usepackage[pdfpagemode=UseNone,pdfstartview=FitH,colorlinks=true,linkcolor=blue,urlcolor=blue,anchorcolor=blue,citecolor=blue]{hyperref}

\usepackage{cancel,xcolor}
\usepackage{indentfirst}
\usepackage{CJKulem}
\usepackage{soul,xcolor}
\setstcolor{red}
\usepackage{ulem}
\usepackage{cancel}

\def\sig{{\mbox{\boldmath{$\sigma$}}}}

\begin{document}

%%%%%%%%%%%%%%%%%%%%%%%%%%%%%%%%%%%%%%%%%%%%%%%%%
%%%%%%%%%%%%%%%%%%%%%%%%%%%%%%%%%%%%%%%%%%%%%%%%%

\title{Magnetizing weak links by time-dependent spin-orbit interactions:\\
momentum conserving and non-conserving processes}
\author{Debashree Chowdhury}
\email{debashreephys@gmail.com}
\affiliation{Centre for Nanotechnology, IIT Roorkee, Roorkee, Uttarakhand  247667, India }

\author{O. Entin-Wohlman}
\email{orawohlman@gmail.com}
\affiliation{School of Physics and Astronomy, Tel Aviv University, Tel Aviv 6997801, Israel}

\author{A. Aharony}
\affiliation{School of Physics and Astronomy, Tel Aviv University, Tel Aviv 6997801, Israel}

\author{R. I. Shekhter}
\affiliation{Department of Physics, University of Gothenburg, SE-412 96 G\"{o}teborg, Sweden}

\author{M. Jonson}
\affiliation{Department of Physics, University of Gothenburg, SE-412 96 G\"{o}teborg, Sweden}

%%%%%%%%%%%%%%%%%%%%%%%%%%%%%%%%%%%%%%%%%%%%%%%%%
%%%%%%%%%%%%%%%%%%%%%%%%%%%%%%%%%%%%%%%%%%%%%%%%%

\begin{abstract}
Rashba spin-orbit interactions generated by time-dependent electric fields acting on weak links (that couple together non-magnetic macroscopic leads) can magnetize the junction. The Rashba spin-orbit interaction that affects the spins of electrons tunneling through the weak links changes their momentum concomitantly. We establish the connection between the magnetization flux induced by processes that conserve momentum and the magnetization created by tunneling events that do not.
Control of the induced magnetization can be achieved by tuning the polarization of the AC electric field responsible for the spin-orbit Rashba interaction (e.g., from being circular to linear), by changing the applied bias voltage, and by varying the degree of a gate voltage-induced asymmetry of the device.  
 
 \end{abstract}
%%%%%%%%%%%%%%%%%%%%%%%%%%%%%%%%%%%%%%%%%%%%%%%%%
%%%%%%%%%%%%%%%%%%%%%%%%%%%%%%%%%%%%%%%%%%%%%%%%%

\date{\today}
\maketitle
%%%%%%%%%%%%%%%%%%%%%%%%%%%%%%%%%%%%%%%%%%%%%%%%%
%%%%%%%%%%%%%%%%%%%%%%%%%%%%%%%%%%%%%%%%%%%%%%%%%

%%%%%%%%%%%%%%%%%%%%%%%%%%%%%%%%%%%%%%%%%%%%%%%%%
%%%%%%%%%%%%%%%%%%%%%%%%%%%%%%%%%%%%%%%%%%%%%%%%%

%\LARGE

\section{introduction}
\label{Intro}

%%%%%%%%%%%%%%%%%%%%%%%%%%%%%%%%%%%%%%%%%%%%%%%%%
%%%%%%%%%%%%%%%%%%%%%%%%%%%%%%%%%%%%%%%%%%%%%%%%%

Tunneling transport of electrons in nanodevices offers an important
tool for the accumulation and control of electric charge in nanometer-sized conductors. Pronounced 
mesoscopic phenomena, such as resonant tunneling of electrons and single-electron tunneling, 
make this control achievable by electrostatic means \cite{Ferry2012, Kastner1992}. 
Spin is another fundamental property of electrons that can be functionalized in tunneling devices 
using spin-dependent tunneling, paving the way for harvesting spintronic phenomena on the 
nanometer-length scale \cite{Fert1990}. One example, which is the focus of this work, is the 
possibility of accumulating a controllable amount of magnetization in quantum single-dot junctions.

%%%%%%%%%%%%%%%%%%%%%%%%%%%%%%%%%%%%%%%%%%%%%%%%%
%%%%%%%%%%%%%%%%%%%%%%%%%%%%%%%%%%%%%%%%%%%%%%%%%

An obvious way to involve the spin of electrons in tunneling transport is to use magnetic materials in hybrid nanodevices. 
While this approach allows for magnetic control of spintronic phenomena, it has a major disadvantage compared to  
the electrostatic means of controlling the charge alluded to above. This is due to practical difficulties in spatially localizing the magnetic field on a nanometer-length scale.
A non-magnetic coupling to the electron spin degree of freedom could alleviate this problem.

%%%%%%%%%%%%%%%%%%%%%%%%%%%%%%%%%%%%%%%%%%%%%%%%%
%%%%%%%%%%%%%%%%%%%%%%%%%%%%%%%%%%%%%%%%%%%%%%%%%

A non-magnetic option of tuning spin-dependent 
phenomena would be to let the spin-orbit interaction (SOI) influence the tunneling of electrons through weak links. Another possibility is using suspended nanowires. These have been shown \cite{Shekhter2013} to provide mechanically controlled coherent mixing or splitting of the spin states of transmitted electrons caused by the Rashba  \cite{Rashba1960} spin-orbit interaction. The sensitivity of this interaction to mechanical bending makes the wire a tunable nano-electromechanical weak link between bulky reservoirs.

%%%%%%%%%%%%%%%%%%%%%%%%%%%%%%%%%%%%%%%%%%%%%%%%%
%%%%%%%%%%%%%%%%%%%%%%%%%%%%%%%%%%%%%%%%%%%%%%%%%

The Rashba spin-orbit interaction, which couples electrons' momenta with their spins, indeed affects electron tunneling through quasi-one-dimensional weak links. 
When generated by an external electric field \cite{Scherubl2016}, it would make it possible for the electronic spin to couple to an electric field. Put another way, the magnitude of an electronic property arising from the interfacial
breaking of inversion symmetry  can be
modulated by applying an external electric field \cite{Caviglia2010}.

Spin polarization of the electrons, occurring as electrons tunnel through an SOI-active weak link, is expected to allow for an electrically generated magnetization of a small-sized non-magnetic device. 
However, a constraint originating from the time-reversal symmetry of the SOI prohibits any effect of it on the two-terminal tunneling of electrons in non-superconducting devices \cite{Bardarson2008}. The role of the spin-orbit interaction in electron transport can be revived by breaking the time-reversal symmetry of this interaction
using an AC electric field. Such a field, rotating with frequency $\Omega$ in a plane perpendicular to the direction of the cylindrical wire forming the junction, does produce a Rashba spin-orbit interaction that does not obey time-reversal symmetry.

%%%%%%%%%%%%%%%%%%%%%%%%%%%%%%%%%%%%%%%%%%%%%%%%%
%%%%%%%%%%%%%%%%%%%%%%%%%%%%%%%%%%%%%%%%%%%%%%%%%

Spin splitting induced by temporal shape modulations of various parts of junctions, i.e., ``pumping", has been proposed in several papers,  with or without additional application of DC or AC magnetic fields, see, for example, Refs. \cite{Sharma2003, Watson2003, Governale2003, Avishai2010, Fajardo2017,Hernangomez2020}. More recent publications study spin pumping into an anisotropic Dirac electron system, generated by shining microwave irradiation on an adjacent ferromagnetic insulator \cite{Funato2022}, or propose to exploit surface states of three-dimensional topological insulators: the spin-momentum locking induces surface spin-density accumulations due to applied electric fields \cite{Asgharpour2020}.

%%%%%%%%%%%%%%%%%%%%%%%%%%%%%%%%%%%%%%%%%%%%%%%%%
%%%%%%%%%%%%%%%%%%%%%%%%%%%%%%%%%%%%%%%%%%%%%%%%%

\begin{figure}
\includegraphics[width=0.43\textwidth]{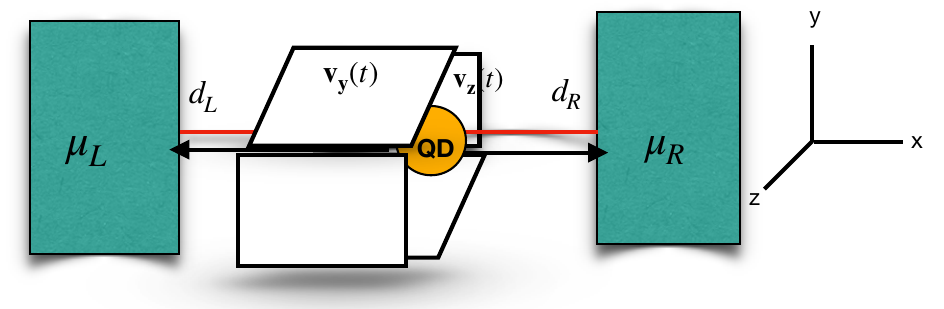}
\caption{(color online) Schematic plot of the device: a single-level (of energy $\varepsilon^{}_{d}$) quantum dot is attached by two weak links (of lengths $d^{}_{L, R}$) to two electron reservoirs, denoted $L$ and $R$, with chemical potentials $\mu^{}_{L,R}$, respectively. The 
 rotating electric fields, produced by the potentials $v^{}_{y,z}(t)$, induce spin-dependent tunneling through the links. }
\label{sys}
\end{figure}

A rotating electric field inducing spin-orbit coupling can result
from two external fields along perpendicular directions, which
are normal to the cylindrical wire (lying along $\hat{\bf x}$), see Fig.~\ref{sys}.
%
%Such fields, rotating with frequency $\Omega$ (using $\hbar=1$ units), are represented by
%\begin{align}
%\hat{\bf n}(t)=\{0,\gamma\sin(\Omega t),-\cos(\Omega t)\}/U(t)\ ,
%\label{n}
%\end{align}
%where the normalization is
%\begin{align}
%U(t)=\sqrt{\cos^{2}(\Omega t)+\gamma^{2}\sin^{2}(\Omega t)}\ ,
%\label{nn}
%\end{align}
%and $\gamma$ measures the amount of ellipticity: $\gamma=0$ corresponds to a linearly-polarized field, while $\gamma=1$ describes a
%circularly-polarized one.
%
Once the spin-orbit Rashba interaction is established in the weak links, the tunneling amplitudes governing them are each augmented by 
a phase factor, $e^{i\varphi_{\rm AC}}$, known as the Aharonov-Casher phase factor.

%%%%%%%%%%%%%%%%%%%%%%%%%%%%%%%%%%%%	
%%%%%%%%%%%%%%%%%%%%%%%%%%%%%%%%%%%%

The Aharonov-Casher phase, $\varphi_{\rm AC}$, is due to the interaction between the
magnetic moment (spin) of an electron and its orbital motion in an electric field \cite{AC1984}. (It is dual to the perhaps more well-known Aharonov-Bohm phase
that results from the movement of a charged particle --- such as an electron --- in a magnetic field.) 

To understand the physics behind the Aharonov-Casher phase, it is useful to consider the classical
analog where a particle carrying magnetic moment ${\boldsymbol \mu}$ moves along a certain trajectory in space
\cite{LTPreview}. 
If an electric field $\bf E$ is switched on, the magnetic moment will start to precess, which can be understood as follows. 
For simplicity, let the electric field be due to a static charge $q$. Then, in the rest frame of the moving particle, it sees a
current due to the charge, which in this frame is moving. 
The magnetic moment will interact with the magnetic field $\bf B$ generated by this current 
and start to precess. 
The interaction energy is given by the usual Zeeman term,
\begin{align}
U = - {\boldsymbol \mu} \cdot {\bf B} = - \frac{1}{2mc^2}\,{\boldsymbol \mu} \cdot
\left( {\bf E} \times {\bf p} \right) \,,
\label{eq1}
\end{align}
where $m$ is the mass of the classical particle and $\bf p$ its momentum.
The generalization of Eq.~(\ref{eq1}) to the case of electrons needs a quantum mechanical and 
relativistic formulation. 
%(since the spin is a relativistic property of electrons). 
The result 
%of what we now call the spin-orbit interaction energy 
is the same 
but with the classical magnetic moment $\boldsymbol \mu$ replaced by the magnetic momentum
operator $-\mu^{}_B {\boldsymbol \sigma}$, where $\mu^{}_B=e\hbar/(2m)$ is the Bohr magneton and 
$\boldsymbol \sigma$ is the vector of Pauli matrices operating on spin space. 
From Eq.~(\ref{eq1}) it is obvious that one can define an effective spin-orbit magnetic field \cite{Jaber},
\begin{align}
{\bf B}_{\rm so} =  \frac{1}{mc^2}
\left( {\bf E} \times {\bf p} \right) \,,
\label{eq2}
\end{align}
which is perpendicular to both the electric field and the momentum of the particle. 
%\footnotetext{
%{\color{blue} [MJ: Maybe the simplest way to make the footnote (colored in red below) visible is to use "backslash cite" ? Or , maybe better, we could include this comment in the text (in which case we should only refer to \cite{Jaber} once.]}
%{\color{magenta} It is to be noted here that Eq. (2) is derived for the quantum spin $1/2$ particles (electrons). Once we replace $\mu_{B}$ in Eq. (2), there is a difference of a factor of $1/2$ from Eq. (1). This is due to the Thomas factor, which is the correction of $1/2$ in the spin-orbit coupling term of the Hamiltonian due to relativistic effects. This factor is included in theoretical calculations to match the experimental observations regarding the fine structure \cite{Jaber}.}
%}
%\color{blue}
%An alternative way to write the footnote (somewhat shorter) might be: \\
Note that Eq. (\ref{eq2}) is derived for electrons, which are quantum particles with spin 1/2. 
This leads to a difference compared to Eq. (\ref{eq1}) by the Thomas factor 1/2, due to a relativistic correction to the spin-orbit coupling  \cite{Jaber}.

%{\color{magenta} [MJ: The new text is taken from Ref. \onlinecite{LTPreview}. 
%I have rewritten the above two equations using SI units, which have added a factor of $c$ to the
%denominators. I am suspicious of the factor 2 in the denominator of Eq. (1), but maybe it is correct.]}

Quantum mechanically, one has to consider an %incident 
electron propagating through the nanowire in Fig.~1 as a plane wave attached to a spinor.
%initially 
%describing the spin projection on the electric field as ``up" as it enters the wire. 
%As the electron propagates along the channel
%its state splits into a linear combination of the spin-up and spin-down states with coefficients determined
%by the precession angle $\varphi$ in the semiclassical picture.
%In quantum mechanics, the phase $\varphi$ is significant in a different sense, called in that context
%the  Aharonov-Casher phase $\varphi_{\rm AC}$. 
It is easy to show that the wave function propagating along the channel accumulates an extra phase
 factor,
\begin{align}
 \varphi_{\rm AC} = - \frac{e}{2mc^2} \int ( {\boldsymbol \sigma} \times {\bf E}) \cdot d{\bf r}.
\label{eq3}
\end{align}
%which for a classical magnetic moment is exactly  the ``exit angle" $\varphi$
%introduced above.
We arrive at this result by noting that the momentum of 
an electron that enters the spin-orbit active channel has to shift, ${\bf p} \to {\bf p} + \delta{\bf p}$, 
to compensate for the spin-orbit interaction energy of Eq.~(\ref{eq1}) and preserve the total energy.
The corresponding extra phase $\int \delta {\bf p} \cdot d{\bf r}$, the Aharonov-Casher phase, is given by Eq.~(\ref{eq3}).

We conclude from the above digression that the only effect of the spin-orbit interaction is to add an 
Aharonov-Casher phase factor exp(i$\varphi_{\rm AC}$) to the transmission amplitude,
\begin{align}
&\oJ^{}_{L}(t)=J^{}_{L,0} e^{{\it i}\varphi_{\rm AC}},
%\exp\Big [ik^{}_{\rm so} d^{}_{L}\sig\cdot[\hat{\bf y}\cos(\Omega t)-\hat{\bf z}\sin(\Omega t)]\Big ]\nonumber\\&
%\equiv\cos(k^{}_{\rm so}d_{L})+i\sin(k^{}_{\rm so}d_{L})[e^{i\Omega t}{\sigma}^{}_{-}+e^{-i\Omega t}{\sigma}^{}_{+}],
\label{Jtun}
\end{align}
rendering the tunneling
to be accompanied by spin flips. 
%This phase factor, arising from the Aharonov-Casher effect \cite{AC1984} modifies the tunneling amplitude

%Such fields,
We consider an electric field  of amplitude $E_0$ directed 
along a unit vector $\hat{\bf n}(t)$ that rotates with angular frequency  $\Omega$ in the $yz$-plane 
%(using units where $\hbar=1$)
%, are represented by
\begin{align}
{\bf E}(t)=E_0 \hat{\bf n}(t); \,\, \hat{\bf n}(t)=\frac{\{0,\gamma\sin(\Omega t),-\cos(\Omega t)\}}{U(t)}\ .
\label{n}
\end{align}
Here the normalization is
\begin{align}
U(t)=\sqrt{\cos^{2}(\Omega t)+\gamma^{2}\sin^{2}(\Omega t)}\ ,
\label{nn}
\end{align}
and $\gamma$ measures the amount of ellipticity: $\gamma=0$ corresponds to a linearly-polarized field, while $\gamma=1$ describes a
circularly-polarized one.

%: for 
For a circularly polarized electric field ($\gamma=1$), using matrix notations, one has
 \cite{Shahbazyan1994}
 \begin{align}
&\frac{\oJ^{}_{L}(t)}{J^{}_{L,0}}=\exp\Big [ik^{}_{\rm so} d^{}_{L}\sig\cdot[\hat{\bf y}\cos(\Omega t)+\hat{\bf z}\sin(\Omega t)]\Big ]\nonumber\\&
\equiv\cos(k^{}_{\rm so}d_{L})+i\sin(k^{}_{\rm so}d_{L})[e^{i\Omega t}{\sigma}^{}_{-}+e^{-i\Omega t}{\sigma}^{}_{+}]\ ,
\label{Jtun}
\end{align}
where
\begin{align}
\sigma^{}_{\mp}=(\sigma^{}_{y}\mp i\sigma^{}_{z})/2\ .
\label{sigmp}
\end{align}
In Eq.~(\ref{Jtun}), $k_{\rm so}=k_{\rm so}(E_0)$ is the strength of the spin-orbit coupling \cite{SOIcomment} induced by the rotating field 
%(in momentum units, with $\hbar=1$) 
 (in units of inverse length) and $d^{}_{L}$ ($d^{}_{R}$) is the length of the weak link connecting the dot to the left (right) terminal. 
%where $ \sigma^{}_{\mp}=(\sigma^{}_{y}\mp i\gamma\sigma^{}_{z})/(2U(t))$.
%In this representation  the strength of the electric field is ``translated" into the Rashba spin-orbit parameter, $k_{\rm so}$ (in momentum units). In Eq. (\ref{ACf}), $d$ is the length of a segment lying parallel to $\hat{\bf x}$ along which the phase is accumulated, and $\sig=\{\sigma^{}_{x},\sigma^{}_{y},\sigma^{}_{z}\}$ is the vector of the Pauli matrices.  

%%%%%%%%%%%%%%%%%%%%%%%%%%%%%%%%%%%%	
%%%%%%%%%%%%%%%%%%%%%%%%%%%%%%%%%%%%

To understand qualitatively the results presented in this paper, it is instructive to focus on the limit of weak SOI coupling, where the Aharonov-Casher phase factor can be expanded in powers of $k_{\rm so}^{}d_{L}^{}$. To the second order in the spin-orbit coupling,
\begin{align}
\exp[ik^{}_{\rm so}d_{L}^{}\Big(\hat{\bf x}&\times\hat{\bf n}(t)\Big) \cdot\sig]\approx 1-(k^{}_{\rm so}d_{L}^{})^{2}/2\nonumber\\
&+i(k^{}_{\rm so}d_{L}^{})[\sigma^{}_{-}e^{i\Omega t}+\sigma^{}_{+}e^{-i\Omega t}]\ .
\label{ACexp}
\end{align}
Note that $\sigma^{}_{\mp}$ in Eq.~(\ref{sigmp}) are the `spin-flip' operators
that increase $(\sigma^{}_{+})$ or decrease $(\sigma^{}_{-})$ the eigenvalue of $\sigma^{}_{x}$ by
one unit when they act on an eigenfunction of $\sigma^{}_{x}$.
Equation (\ref{ACexp}) implies that electrons tunneling from the
reservoirs to the quantum dot may flip their spin. Moreover, such spin flips are accompanied by the emission or absorption of a quantum $\Omega$ of the time-dependent spin-orbit interaction. The spin flips are correlated
with these processes in such a way that a spin-up transition is
accompanied by the emission of an energy quantum, while a
spin-down flip is induced by the absorption of it (for $\Omega>0$).
The Aharonov-Casher factors (\ref{ACexp}) are written for a circularly polarized electric field. Still, the correlation between spin
flips and absorption/emission of quanta persists even for an elliptically polarized field but to a lesser degree.  It disappears
only for linear polarization, for which $\sigma_{+}=\sigma^{}_{-}=\sigma^{}_{y}$.
The optimal possibility for creating magnetizations is the circularly polarized field ($\gamma=1$), considered in this study.

%{\color{blue} MJ: Delete the following paragraph (blue font) and replace it with the text (in magneta) that follows:

%Time-dependent Rashba interactions can generate DC charge currents even in the absence of a bias voltage \cite{O2020}, in addition to transverse components of the spin-polarized currents, which rotate in the plane perpendicular to the weak links. These transverse
%components vanish upon averaging over the field's periodicity and thus would not appear in the “standard” spin-pumping approach. The DC charge {\color{blue} {\it current}} is carried by electrons whose spins are polarized parallel to the weak links; it disappears for a geometrically symmetric junction.}

Even in the absence of a bias voltage, time-dependent AC Rashba interactions can generate DC charge 
currents through a non-symmetric junction such as the one shown in Fig.~1 as well as generate a voltage drop over the 
junction if the junction is part of an open circuit \cite{O2020}. The purpose of the present paper is to show that a time-dependent Rashba interaction can also magnetize the junction.

%%%%%%%%%%%%%%%%%%%%%%%%%%%%%%%%%%%%%%%%%%%%%%%%%
%%%%%%%%%%%%%%%%%%%%%%%%%%%%%%%%%%%%%%%%%%%%%%%%%

%%%%%%%%%%%%%%%%%%%%%%%%%%%%%%%%%%%%	
%%%%%%%%%%%%%%%%%%%%%%%%%%%%%%%%%%%%

\subsection{Formulation of the problem}
\label{prob}

%%%%%%%%%%%%%%%%%%%%%%%%%%%%%%%%%%%%%%%%%%%%%%%%%%%%%%%%%%%

Our calculation is based on the tunneling Hamiltonian for the single-dot junction,  
\begin{align}
{\cal H}(t)={\cal H}^{}_{L}+{\cal H}^{}_{R}+{\cal H}^{}_{\rm dot}+{\cal H}^{}_{\rm tun}(t)\ ,
\label{H}
\end{align}
that consists of the  Hamiltonians of the macroscopic leads and the dot  
\begin{align}
&{\cal H}^{}_{L}+{\cal H}^{}_{R}+{\cal H}^{}_{\rm dot}\nonumber\\
&=\sum_{{\bf k},\sigma}\epsilon^{}_{k}c^{\dagger}_{{\bf k}\sigma}c^{}_{{\bf k}\sigma}+
\sum_{{\bf p},\sigma}\epsilon^{}_{p}c^{\dagger}_{{\bf p}\sigma}c^{}_{{\bf p}\sigma}+\epsilon^{}_{d}\sum_{\sigma}c^{\dagger}_{d\sigma}c^{}_{d\sigma}\ .
\label{Hs}
\end{align}
The left (right) lead is described by the creation operators $c^{\dagger}_{{\bf k}\sigma}$ and energy $\epsilon^{}_{k}$ ($c^{\dagger}_{{\bf p}\sigma}$ and $\epsilon^{}_{p}$), the dot by the creation operator $c^{\dagger}_{d\sigma}$ and site energy $\epsilon^{}_{d}$.  The tunneling Hamiltonian in the weak links is  
\begin{align}
{\cal H}^{}_{\rm tun}(t)&=
\sum_{{\bf k}}\sum_{\sigma,\sigma'}[c^{\dagger}_{{\bf k}\sigma}\oJ^{}_{L,\sigma\sigma'}(t)c^{}_{d\sigma'}+{\rm H.c.}]\nonumber\\
&+
\sum_{{\bf p}}\sum_{\sigma,\sigma'}[c^{\dagger}_{{\bf p}\sigma}\oJ^{}_{R,\sigma\sigma'}(t)c^{}_{d\sigma'}+{\rm H.c.}]
\ .
\label{Htun}
\end{align}
For a Rashba interaction generated by a circularly polarized electric field, the tunneling amplitude in the left link (of length $d_{L}^{}$) is given in 
Eq.~(\ref{Jtun}),
with $\sigma^{}_{\mp}$ given in Eq.~(\ref{sigmp}).
The tunnel coupling in the right link is obtained upon changing $J^{}_{L,0}$, the bare (SOI independent) tunneling amplitude, into $J^{}_{R,0}$, and $d_{L}^{}$ into $-d^{}_{R}$, see Fig.~\ref{sys}.

%%%%%%%%%%%%%%%%%%%%%%%%%%%%%%%%%%%%%%%%%%%%%%%%%%%%%%%%%%%
%%%%%%%%%%%%%%%%%%%%%%%%%%%%%%%%%%%%%%%%%%%%%%%%%%%%%%%%%%%

The magnetizations created on the dot, ${\bf M}^{}_{d}(t)$,  and in the reservoirs, ${\bf M}^{}_{L,R}(t)$, are {\it a priori} time-dependent. In units of $-g\mu^{}_{\rm B}/2$ ($g$ is the $g-$factor of the electron and $\mu^{}_{\rm B}$ is the Bohr magneton) these are 
\begin{align}
{\bf M}^{}_{d}(t)=\sum_{\sigma,\sigma'}\langle c^{\dagger}_{d\sigma}(t)\sig^{}_{\sigma\sigma'}c^{}_{d\sigma'}(t)\rangle\ ,
\label{dot}
\end{align}
and 
\begin{align}
{\bf M}^{}_{L}(t)&=\sum_{{\bf k},{\bf k}'}\sum_{\sigma,\sigma'}\langle c^{\dagger}_{{\bf k}\sigma}(t)\sig^{}_{\sigma\sigma'}c^{}_{{\bf k}'\sigma'}(t)\rangle%=\sum_{{\bf k},{\bf k}'}{\rm Tr}\{-iG^{<}_{{\bf k}{\bf k}'}(t,t)\sig\}
\ ,\nonumber\\
{\bf M}^{}_{R}(t)&=\sum_{{\bf p},{\bf p}'}\sum_{\sigma,\sigma'}\langle c^{\dagger}_{{\bf p}\sigma}(t)\sig^{}_{\sigma\sigma'}c^{}_{{\bf p}'\sigma'}(t)\rangle%=\sum_{{\bf p},{\bf p}'}{\rm Tr}\{-iG^{<}_{{\bf p}{\bf p}'}(t,t)\sig\}
\ ,
\label{lead}
\end{align}
where the angular brackets indicate quantum-mechanical averaging with the Hamiltonian (\ref{H}). 
The time derivatives of ${\bf M}_{d}^{}(t)$ and ${\bf M}^{}_{L,R}(t)$ give the magnetization (spin) fluxes. 

%%%%%%%%%%%%%%%%%%%%%%%%%%%%%%%%%%%%%%%%%%%%%%%%%%%%%%%%%%%
%%%%%%%%%%%%%%%%%%%%%%%%%%%%%%%%%%%%%%%%%%%%%%%%%%%%%%%%%%%

The magnetizations and their fluxes are conveniently expressed in terms of the time-dependent Keldysh Green's functions, in particular, the lesser one $G^{<} _{\alpha,\alpha'}(t,t')=i\langle c^{\dagger}_{\alpha'}(t')c^{}_{\alpha}(t)\rangle$, where $\alpha$ represents the degrees of freedom of the electrons \cite{Jauho1994, Haug2008}. Thus, the magnetization formed on the dot is
\begin{align}
&{\bf M}^{}_{d}(t)={\rm Tr}\{-iG^{<}_{dd}(t,t)\sig\}\ , 
\label{dotM}
\end{align}
where $G^{<}_{dd}$ (the Green's function on the dot) is a (2$\times$2) matrix in spin space, and the ones generated in the leads are
\begin{align}
{\bf M}^{}_{L(R)}(t)=\sum_{{\bf k}({\bf p}),{\bf k}'({\bf p}')}{\rm Tr}\{-iG^{<}_{{\bf k}({\bf p}){\bf k}'({\bf p}')}(t,t)\sig\}\ . 
\label{ML}
\end{align}
These Green's functions are worked out in Appendix \ref{GF}.

%%%%%%%%%%%%%%%%%%%%%%%%%%%%%%%%%%%%%%%%%%%%%%%%%%%%%%%%%%%
%%%%%%%%%%%%%%%%%%%%%%%%%%%%%%%%%%%%%%%%%%%%%%%%%%%%%%%%%%%

Neither ${\bf M}^{}_{L}(t)$ nor ${\bf M}^{}_{R}(t)$ are diagonal in momentum space: the Rashba interaction couples the spin of the tunneling electrons with their momentum, and the tunneling  Hamiltonian (\ref{Htun}) destroys the momentum conservation embedded in the (decoupled) leads' Hamiltonians ${\cal H}_{L,R}^{}$. Separating the contributions to the lead magnetizations into momentum-conserving ones, resulting from $G^{<}_{\bf kk}$ ($G^{<}_{\bf pp}$), and non-conserving ones, $G^{<}_{{\bf kk}'}$ ($G^{<}_{{\bf pp}'}$), with ${\bf k}\neq{\bf k}'$ (${\bf p}\neq{\bf p}'$), we derive in Appendix \ref{GF} 
 (using units where $\hbar=1$), the (exact) matrix relation
\begin{align}
\frac{d}{dt}\sum_{{\bf k},{\bf k}'}G^{<}_{{\bf k}{\bf k}'}(t,t)=&\frac{d}{dt}\sum_{{\bf k}}G^{<}_{{\bf k}{\bf k}}(t,t)\nonumber\\
&+i\sum_{{\bf k},{\bf k}'}(\epsilon^{}_{{\bf k}'}-\epsilon^{}_{{\bf k}})G^{<}_{{\bf k}{\bf k}'}(t,t)\ ,
\label{derGkk}
\end{align}
(and a similar one for the right lead). Below, we refer to the magnetization flux due to the equal-momentum processes (aka   momentum conserving) as ${\bf M}^{\rm con}_{L(R)}$ and denote by 
${\bf M}^{\rm non-con}_{L(R)}$ the magnetization due to  ${\bf k}\neq {\bf k}'$ (${\bf p}\neq{\bf p}'$), i.e., the momentum non-conserving ones.

%%%%%%%%%%%%%%%%%%%%%%%%%%%%%%%%%%%%%%%%%%%%%%%%%%%%%%%%%%%
%%%%%%%%%%%%%%%%%%%%%%%%%%%%%%%%%%%%%%%%%%%%%%%%%%%%%%%%%%%

Our calculation utilizes the approximation termed the `wide-band limit' \cite{Jauho1994, Haug2008} (see Sec. \ref{WBA}). In this framework,  there arises a difficulty in computing the magnetization due to momentum-conserving processes, explained in Appendix \ref{GF},  in the text around Eqs. (\ref{con1}) and (\ref{con2}). This difficulty disappears from the expression for the {\it magnetization flux}, of the momentum-conserving processes, $\dot{\bf M}^{\rm con}_{L,R}(t)$. We derive in Appendix \ref{GF} a relation between ${\bf M}^{\rm non-con}_{L,R}(t)$ and $\dot{\bf M}^{\rm con}_{L,R}(t)$, 
\begin{align}
\dot{\bf M}^{\rm con}_{L(R)}(t)={\bf M}^{\rm non-con}_{L(R)}(t)/(\pi{\cal N}^{}_{L(R)})\ ,
\label{MconMnon}
\end{align}
where ${\cal N}^{}_{L(R)}$ is the density of states at the Fermi energy in the left (right) lead (of dimension [energy]$^{-1}$). 
This relation shows that the momentum-conserving processes' contribution to the magnetization flux considerably exceeds
the magnetization resulting from the momentum non-conserving terms due to the presence of the density of states (${\cal N}_{L,R})$, which makes the contribution of the momentum non-conserving processes rather small. However, it is important to note that ${\bf M}^{\rm non-con}_{L(R)}$ includes a time-independent component along the weak link direction $\hat{\bf x}$,  and thus survives averaging over the electric field's periodicity. Such a time-independent $\hat{\bf x}$ component also appears in the magnetization created on the dot,  ${\bf M}^{}_{d}(t)$  \cite{OEW2020}.

%%%%%%%%%%%%%%%%%%%%%%%%%%%%%%%%%%%%%%%%%%%%%%%%%%%%%%%%%%%
%%%%%%%%%%%%%%%%%%%%%%%%%%%%%%%%%%%%%%%%%%%%%%%%%%%%%%%%%%%

\subsection{The wide-band limit}

%%%%%%%%%%%%%%%%%%%%%%%%%%%%%%%%%%%%%%%%%%%%%%%%%%%%%%%%%%%

\label{WBA}

%%%%%%%%%%%%%%%%%%%%%%%%%%%%%%%%%%%%%%%%%%%%%%%%%%%%%%%%%%%

The detailed derivation of Keldysh Green's functions is presented in Appendix \ref{App}. Here, we outline the main 
approximation involved in this calculation, which is the so-called ``wide-band approximation". This
approximation ignores the energy dependence of the electron density of states in the electrodes and is good if
the electrons there form a degenerate electron gas, i.e., if
$k_B T \ll \mu_{L,R} \, $ \cite{temperature},

The Keldysh Green's function $G_{dd}$ is built from the decoupled dot Green's function (denoted $g^{}_{d}$), and the self-energy $\Sigma$ created on the dot due to its coupling with the two reservoirs.
Quite generally, the self-energy  $\Sigma=\Sigma^{}_{L}+\Sigma^{}_{R}$, where
$\Sigma^{}_{L}$, resulting from the coupling with the left reservoir, is given by
\begin{align}
\Sigma^{}_{L}(t,t')=\oJ^{\dagger}_{L}(t)\oJ^{}_{L}(t')\sum_{\bf k}g^{}_{\bf k}(t-t')\ ,
\label{SigmaL}
\end{align}
(and similarly for $\Sigma^{}_{R}$).
Here, $g^{}_{{\bf k}}(t-t')$ [described by ${\cal H}_{L}$ in Eq.~(\ref{Hs})] is the Green's function of the decoupled left lead.
For time-dependent tunneling amplitudes $\oJ^{}_{L}(t)$, the expression in Eq.~(\ref{SigmaL}) is based on a possible factorization \cite{Jauho1994, Haug2008}
of the momentum and time dependencies of the tunneling couplings [see Eq.~(\ref{Jtun}), where $J^{}_{L,0}$ is taken at the Fermi momentum]. This is termed the ``wide-band limit"

%%%%%%%%%%%%%%%%%%%%%%%%%%%%%%%%%%%%%%%%%%%%%%%%%%%%%%%%%%%

Presenting the (retarded and advanced) Green's functions, $g^{r,a}_{\bf k}$, of the uncoupled left lead in terms of their Fourier transforms
\begin{align}
g^{r,a}_{{\bf k}}(t-t')=\int\frac{d\omega}{2\pi}\frac{e^{-i\omega(t-t')}}{\omega-\epsilon^{}_{k}\pm i0^{+}_{}}\ ,
\label{WBL}
\end{align}
one finds that
within  the wide-band approximation \cite{Haug2008}, 
\begin{align}
\sum_{\bf k}g^{r,a}_{\bf k}(t-t')\approx\mp i\pi{\cal N}^{}_{L}\delta(t-t')\  .
\label{del}
\end{align}
This is based on the assumption that--since mainly electrons close to the Fermi energy participate in the tunneling--the dependence of the density of states on the energy may be ignored \cite{Haug2008}. Hence, noting that $\oJ^{}_{L}$ is proportional to a unitary matrix,  the retarded, advanced, and lesser self-energies are
\begin{align}
&\Sigma^{r,a}_{L}(t,t')=\mp i\Gamma^{}_{L}\delta(t-t')\ ,\  
\Gamma^{}_{L}=\pi{\cal N}^{}_{L}|J^{}_{L,0}|^{2}\ ,\nonumber\\
&\Sigma^{<}_{L}(t,t')=i\sum_{{\bf k}}f^{}_{L}(\epsilon^{}_{k})e^{-i\epsilon^{}_{k}(t-t')}\oJ^{\dagger}_{L}(t)\oJ^{}_{L}(t')\ ,
\label{sigral}
\end{align}
$f^{}_{L}(\epsilon^{}_{k})$ being the Fermi distribution in the left lead. Analogous calculation holds for the right lead. The width of the resonance formed on the dot due to the coupling with the leads is $\Gamma=\Gamma^{}_{L}+\Gamma^{}_{R}$.

%%%%%%%%%%%%%%%%%%%%%%%%%%%%%%%%%%%%%%%%%%%%%%%%%%%%%%%%%%%
%%%%%%%%%%%%%%%%%%%%%%%%%%%%%%%%%%%%%%%%%%%%%%%%%%%%%%%%%%%

Our results for the magnetization are presented in Sec. \ref{spinden}; the detailed explicit derivations are relegated to Appendix \ref{App}. 
%%%%%%%%%%%%%%%%%%%%%%%%%%%%%%%%%%%%	
%%%%%%%%%%%%%%%%%%%%%%%%%%%%%%%%%%%%

%\section{The spin densities}
\section{Magnetization}

\label{spinden}

We present the magnetizations in dimensionless units, omitting the factor 
$g\mu^{}_{\rm B}/2$, where %{\st{$g=0.45$ as measured for conduction electrons in GaAs [cite(Zawadzki2008)]},
$g$ is the electron spin $g$-factor.% \st{and $\mu^{}_{\rm B}=e\hbar/(2m)$}
%{\color{magenta} [MJ: I suggest to keep it more general by not giving a specific value for $g$ here.
%The InAs nanowire system studied by Sherubl et al seems more close to what we suggest than GaAs.
%(I deleted the reference to the paper by Zawadski which only seems to be cited here; the Bohr magneton is now defined in the Introduction)]}

The derivation of the magnetization generated on the dot and in the leads is detailed in Appendix \ref{App}. 
For low enough AC frequencies, smaller than the width of the Breit-Wigner resonance formed on the dot ($\Omega<\Gamma$),
the magnetization induced on the dot is
\begin{align}
{\bf M}_{d}&(t)=
 \frac{2\Omega}{\pi}\Big (\Gamma^{}_{L}{\cal D}^{}_{L}
 [\sin^{2}(k^{}_{\rm so}d^{}_{L})\hat{\bf x}-\frac{1}{2}\sin(2k^{}_{\rm so}d^{}_{L})\hat{\bf n}(t)]\nonumber\\
 &+\Gamma^{}_{R}{\cal D}^{}_{R}
 [\sin^{2}(k^{}_{\rm so}d^{}_{R})\hat{\bf x}+\frac{1}{2}\sin(2k^{}_{\rm so}d^{}_{R})\hat{\bf n}(t)]\Big )\ .
 \label{dotMf}
\end{align}
Here, 
\begin{align}
{\cal D}^{}_{L(R)}&=\int d\epsilon^{}_{k(p)}\frac{-\partial f^{}_{L(R)}(\epsilon^{}_{k(p)})/\partial\epsilon^{}_{k(p)}}{(\epsilon^{}_{d}-\epsilon^{}_{k(p)})^{2}+\Gamma^{2}}\nonumber\\
&\approx[(\epsilon^{}_{d}-\mu^{}_{L(R)})^{2}+\Gamma^{2}]^{-1}\ ,
\end{align}
represents the Breit-Wigner resonance formed on the dot due to the coupling with each lead (the second expression pertains to very low temperatures).  The unit vector $\hat{\bf n}(t)$ characterizes the circularly polarized AC electric field, Eq.~(\ref{n}) with $\gamma=1$.
The unit vectors 
$[\sin(k^{}_{\rm so}d^{}_{L})\hat{\bf x}-\cos(k^{}_{\rm so}d^{}_{L})\hat{\bf n}(t)]$ and $ [\sin(k^{}_{\rm so}d^{}_{R})\hat{\bf x}+\cos(k^{}_{\rm so}d^{}_{R})\hat{\bf n}(t)]$
define the orientation of the magnetization on the dot.

%%%%%%%%%%%%%%%%%%%%%%%%%%%%%%%%%%%%%%%%%%%%%%%%%%%%%%%%%%%
%%%%%%%%%%%%%%%%%%%%%%%%%%%%%%%%%%%%%%%%%%%%%%%%%%%%%%%%%%%

For a symmetric ($d_{L}^{}=d^{}_{R}$, $\Gamma^{}_{L}=\Gamma^{}_{R}$), unbiased junction ($\mu_{L}=\mu_{R}$), the magnetization created on the dot is time-independent and directed along the weak links' axis $\hat{\bf x}$. Biasing the junction, $\mu^{}_{L(R)}=\mu\pm eV/2$,  leads to
\begin{align}
&{\cal D}^{}_{L(R)}\approx {\cal D}[1\pm(\epsilon^{}_{d}-\mu) eV{\cal D}]\ ,\nonumber\\
& {\cal D}=[(\epsilon^{}_{d}-\mu)^{2}+\Gamma^{2}]^{-1}\ ,
\end{align}
so that the magnetization induced in a symmetric dot is
\begin{align}
&{\bf M}_{d}(t)=
 (2/\pi)\Gamma\Omega{\cal D}\sin(k^{}_{\rm so}d^{}_{L})\nonumber\\
 &\times[\sin^{}(k^{}_{\rm so}d^{}_{L})\hat{\bf x}-(\epsilon^{}_{d}-\mu) eV{\cal D}\cos(k^{}_{\rm so}d^{}_{L})\hat{\bf n}(t)]\ .
\end{align}
Namely, a bias voltage induces a time-dependent component to the dot magnetization.

%%%%%%%%%%%%%%%%%%%%%%%%%%%%%%%%%%%%%%%%%%%%%%%%%%%%%%%%%%%
%%%%%%%%%%%%%%%%%%%%%%%%%%%%%%%%%%%%%%%%%%%%%%%%%%%%%%%%%%%

The magnetization formed on the leads results from the momentum-conserving and momentum-non-conserving processes. We derive in Appendix \ref{App} the magnetization {\it flux} for the first type, $\dot{\bf M}^{\rm con}_{L(R)}(t)$, for reasons explained around Eqs. (\ref{con1}) and (\ref{con2}), and show that within the wide-band limit, it is related to the magnetization created by the momentum non-conserving processes, ${\bf M}^{\rm non-con}_{L(R)}(t)$, see Eq.~(\ref{MconMnon}).

The magnetization fluxes due to momentum-conserving processes are
\begin{widetext}
\begin{align}
&\dot{\bf M}^{\rm con}_{L}(t)\approx\ (4\Omega\Gamma^{}_{L}\Gamma^{}_{R}/\pi)\Big ({\cal D}^{}_{L}[\sin^{2}(k^{}_{\rm so}d^{}_{L})\hat{\bf x}+\frac{\sin(2k^{}_{\rm so}d^{}_{L})}{2}\hat{\bf n}(t)]
\nonumber\\
&+{\cal D}^{}_{R}\sin(k^{}_{\rm so}d^{}_{R})[\sin(k^{}_{\rm so}[2d^{}_{L}+d^{}_{R}])\hat{\bf x}+\cos(k^{}_{\rm so}[2d^{}_{L}+d^{}_{R}])\hat{\bf n}(t)]\Big )\ ,
\end{align}
and
\begin{align}
&\dot{\bf M}^{\rm con}_{R}(t)\approx (4\Omega\Gamma^{}_{L}\Gamma^{}_{R}/\pi)\Big ({\cal D}^{}_{R}[\sin^{2}(k^{}_{\rm so}d^{}_{R})\hat{\bf x}-\frac{\sin(2k^{}_{\rm so}d^{}_{R})}{2}\hat{\bf n}(t)]
\nonumber\\
&+{\cal D}^{}_{L}\sin(k^{}_{\rm s0}d^{}_{L})[\sin(k^{}_{\rm so}[2d^{}_{R}+d^{}_{L}])\hat{\bf x}-\cos(k^{}_{\rm so}[2d^{}_{R}+d^{}_{L}])\hat{\bf n}(t)]\Big )\ .
\end{align}
The longer lengths $2d^{}_{L}+d^{}_{R}$ and 
 $2d^{}_{R}+d^{}_{L}$ 
 reflect the ``longer" distances the tunneling electron travels from the left (right) to the right (left) reservoir, going through the right (left) link once and through the left (right) link twice. In particular, the sum $\dot{\bf M}^{\rm con}_{L}(t)+\dot{\bf M}^{\rm con}_{R}(t)$, and consequently also the sum
${\bf M}^{\rm non-con}_{L}(t)+{\bf M}^{\rm non-con}_{R}(t)$, are {\it time-indepependent}, producing a constant magnetization, 
\begin{align}
\dot{\bf M}^{\rm con}_{L}+\dot{\bf M}^{\rm con}_{R}\approx (8\Omega\Gamma^{}_{L}\Gamma^{}_{R}/\pi)\sin(k^{}_{\rm so}[d^{}_{L}+d^{}_{R}])&\Big ({\cal D}^{}_{L}\sin(k^{}_{\rm so}d^{}_{L})\cos(k^{}_{\rm so}d^{}_{R})
+{\cal D}^{}_{R}\sin(k^{}_{\rm so}d^{}_{R})\cos(k^{}_{\rm so}d^{}_{L})\Big )\hat{\bf x}\ .
\end{align}
\end{widetext}

%%%%%%%%%%%%%%%%%%%%%%%%%%%%%%%%%%%%%%%%%%%%%%%%%%%%%%%%%%% 
%%%%%%%%%%%%%%%%%%%%%%%%%%%%%%%%%%%%%%%%%%%%%%%%%%%%%%%%%%%

\begin{figure}
\includegraphics[width=0.4\textwidth]{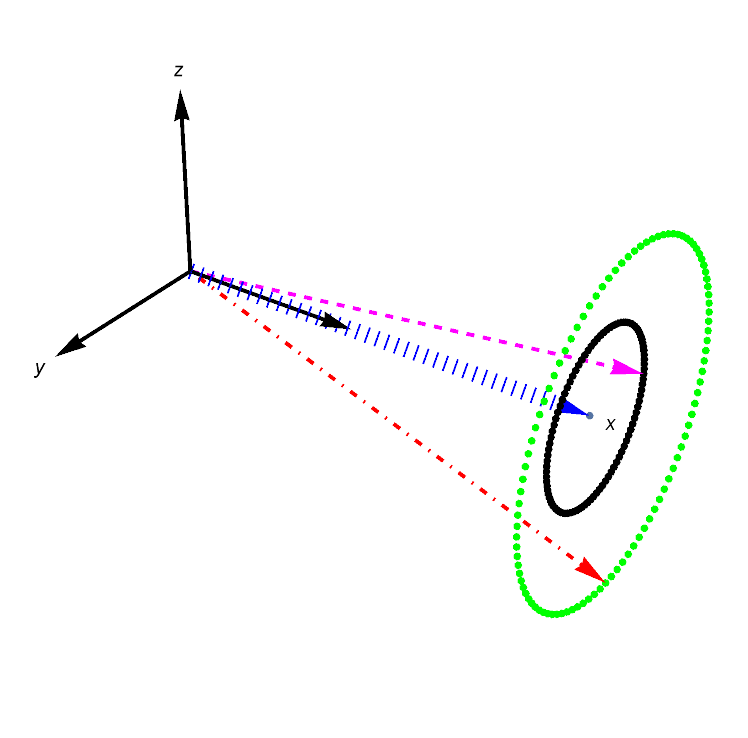}
\caption{
The magnetization vector induced on the dot. The slanted line (blue) vector that coincides with the $\hat{\bf x}$ direction is for $d_{L}=d_{R}$, and is time-independent. The dashed line (magenta) vector lying on the $x-z$ plane is for $d^{}_{R}=1.5 d^{}_{L}$ and $\Omega t=\pi/2$, and the dot-dashed (red) vector in the $x-y$ is for $d^{}_{R}=0.5 d^{}_{L}$ and $\Omega t=\pi$. The circles (black and Green) show the entire rotation for $t=\{0, 2\pi/\Omega\}$.}
\label{fig2}
\end{figure}

%%%%%%%%%%%%%%%%%%%%%%%%%%%%%%%%%%%%%%%%%%%%%%%%%%%%%%%%%%% 
%%%%%%%%%%%%%%%%%%%%%%%%%%%%%%%%%%%%%%%%%%%%%%%%%%%%%%%%%%%

For concreteness, we consider an SOI-active weak link in the form of an InAs
nanowire, for which a value of $k_{\rm so}=1/(100{\rm nm})$ was measured by Scher\"{u}bl {\it et al.} \cite{Scherubl2016}, and more recently in Refs. \cite{Chen2020, Wikacsono2024}. A wire of length equal to 100 nm would
then give $k_{\rm so}d$ of the order 1. One may also choose $\epsilon^{}_{d}-\mu$ to be of the order of $\Gamma$ \cite{Ludovico2016}. The magnitude of $\Gamma$ can be found using typical InAs nanowire values, here taken from Ref. \cite{Dayeh2010}. Exploiting those, we have found  \cite{Jonson2019} that $\Gamma\approx 3 {\rm meV}$. For $\hbar\Omega \approx 0.1$ meV one finds $\Omega=2\pi\times 20$ Ghz, ensuring the calculation to be in the low-frequency regime. As a result, the magnitude of the magnetization created on the dot (in units of $g\mu^{}_{\rm B}/2$) is $\Omega/(\pi\Gamma)\approx 0.1$.
Figure~\ref{fig2} portrays the vector of the cyclic motion ($t=\{0,2\pi/\Omega$\}) of the magnetization formed on the dot, 
for $\Gamma_{L}=\Gamma_{R}=\Gamma/2$, and in the absence of a bias, i.e., for ${\cal D}_{L}={\cal D}^{}_{R}\equiv {\cal D}$. The figure shows 
$\pi{\bf M}^{}_{d}(t)/(\Omega \Gamma {\cal D})\approx 20{\bf M}^{}_{d}(t)$ [see Eq.~(\ref{dotMf})].

%%%%%%%%%%%%%%%%%%%%%%%%%%%%%%%%%%%%%%%%%%%%%%%%%%%%%%%%%%%
%%%%%%%%%%%%%%%%%%%%%%%%%%%%%%%%%%%%%%%%%%%%%%%%%%%%%%%%%%%
The magnetization created,  for instance on the left lead (brought about by the momentum-non-conserving processes)
is much smaller compared to that induced on the dot, 
by the dimensionless factor $\pi\Gamma{\cal N}^{}_{L}$ (assuming for simplicity that $\Gamma_{L}=\Gamma_{R}=\Gamma/2$).
The authors of Ref. \cite{Kwapinski2003} estimated the product $\Gamma {\cal N}$ to be $100^{-1}$.  Then  $\Gamma\approx 3 {\rm meV}$ implies
${\cal N}_{}\approx 1/(0.5 \times 10^{-12}{\rm erg}) $ for the value of $\Gamma {\cal N}$ given in Ref. \cite{Kwapinski2003}. Alternatively,  $\Gamma{\cal N}$ can be estimated by considering two bulky reservoirs coupled by a single weak link. In that case, the tunneling conductance in units of the quantum conductance $G_{0}=e^{2}/(\pi \hbar)$ is 
\begin{align}
G/G^{}_{0}\approx4\pi^{2}|J^{}_{0}|^{2}{\cal N}^{}_{L}{\cal N}^{}_{R}\ .
\end{align}
For typical values of $G/G_{0}$ of InAs nanowires \cite{Scherubl2016, Chuang2012, Dayeh2010}, one obtains a considerably higher estimate,  $\Gamma {\cal N}\approx 10^{4}$, that, in turn, reduces the value of the magnetization generated in the leads by the momentum-non-conserving processes considerably as compared to the one that induced on the dot.
Note that the magnetization {\it flux} in the left link due to the momentum-conserving processes is larger, by a factor of ${\cal N}^{-1}_{L}$.

\begin{figure}
\includegraphics[width=0.3\textwidth]{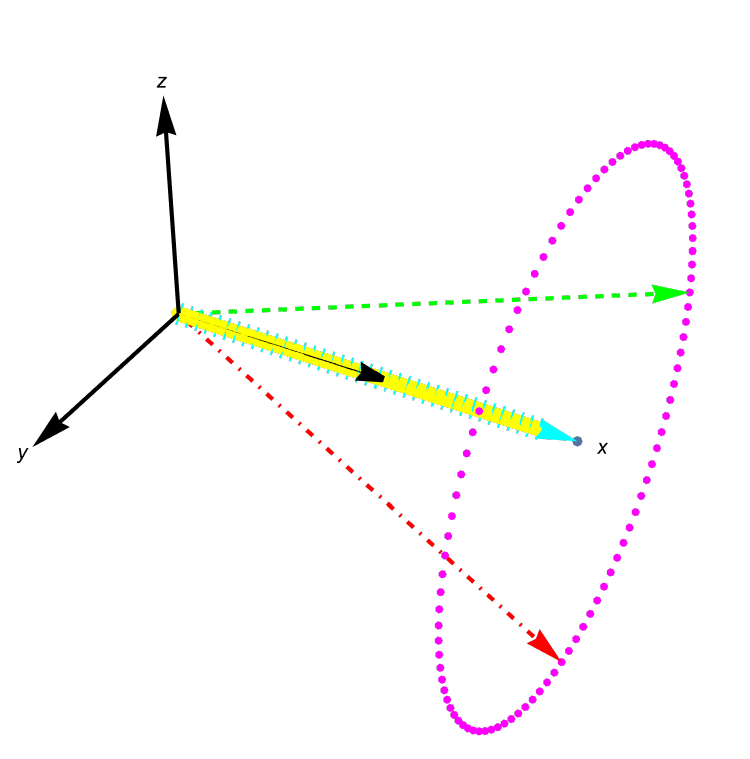}
\caption{(color online) The vectors of the magnetizations formed on the leads, for $d^{}_{L}=d^{}_{R}$.The left lead magnetization has components along $\hat{\bf x}$ (yellow solid line) and along $\hat{\bf z}$ at $\Omega t=\pi/2$, portrayed by the dashed (green) arrow. The dot-dashed (red) arrow shows the magnetization vector in the right lead at $\Omega t=\pi$. The cyan vertical dashed line along $\hat{\bf x}$ shows the $\hat{\bf x}$ component of the right lead magnetization vector. The circle in magenta shows the entire rotation for $t=\{0, 2\pi/\Omega \}$. }
\label{fig3}
\end{figure}

%%%%%%%%%%%%%%%%%%%%%%%%%%%%%%%%%%%%%%%%%%%%%%%%%%%%%%%%%%%
%%%%%%%%%%%%%%%%%%%%%%%%%%%%%%%%%%%%%%%%%%%%%%%%%%%%%%%%%%%
The vectors determining the magnetization flux in the leads due to momentum-conserving processes (or equivalently, the vectors of the magnetizations formed by momentum-non-conserving processes) are depicted in Fig.~\ref{fig3}, for $\Gamma_{L}=\Gamma_{R}=\Gamma/2$, and an unbiased junction, for which ${\cal D}_{L}^{}={\cal D}^{}_{R}={\cal D}$.

%The vector of the lead magnetizations $\frac{M_{L(R)}^{non-con}}{(\Omega/\Gamma)({\cal N}_{L(R)}\Gamma)}[1+(\frac{\epsilon_{d}-\mu}{\Gamma})^{2}]$ are plotted for an unbiased junction in the lower plot. The left lead magnetization has components along x axis (cyan solid arrow along x) and in the x-y plane, which is indicated by the green dashed arrow. Also the right lead magnetization has component along the x axis and in the x-z plane (red dot-dashed arrow). The plot is for $d_{L}=d_{R}.$ The length of the vectors for $M_{L}$ and $M_{R}$ in the x-y and x-z plane have different magnitudes as is evident from the Green and red arrows in the plot.

\section{Conclusions}
\label{conclu}

%%%%%%%%%%%%%%%%%%%%%%%%%%%%%%%%%%%%%%%%%%%%%%%%%%%%%%%%%%%
%%%%%%%%%%%%%%%%%%%%%%%%%%%%%%%%%%%%%%%%%%%%%%%%%%%%%%%%%%%

Perhaps the most conspicuous feature of these results is the appearance of nondissipative (i.e., time-independent) magnetizations along the junction direction $\hat{\bf x}$. Particularly amazing is the nondissipative character of ${\bf M}^{}_{L}+{\bf M}^{}_{R}$, that is solely along the $\hat{\bf x}$ component, independent of whether the junction is symmetric and/or in the presence of a bias voltage. Interestingly, this is not the case for the magnetization formed on the dot--its time dependence disappears only in a symmetric junction.

%%%%%%%%%%%%%%%%%%%%%%%%%%%%%%%%%%%%%%%%%%%%%%%%%%%%%%%%%%%
%%%%%%%%%%%%%%%%%%%%%%%%%%%%%%%%%%%%%%%%%%%%%%%%%%%%%%%%%%%

The Rashba coupling induced by a static electric field does not lead to spin transport in a two-terminal setup. This is due to the time-reversal symmetry of the Aharonov-Casher static phase factor \cite{AC1984}. Previous attempts at generating spin magnetization and transport by static Rashba interactions had to involve the application of current (i.e., a bias voltage) on an {\it asymmetric} junction, see for instance Ref. \onlinecite{Reynoso2024}.
A time-periodic AC electric field breaks the time-reversal symmetry of the Rashba coupling.  As a result,  many inelastic channels are open for the electrons. The inelastic tunneling processes of electrons adeptly produce a net spin population in the two-terminal setups. This effect shows up due to the combined fact that the tunneling electrons can emit or absorb photons, and also due to certain spin flips.  Thus, driving a system with a circularly polarized light-induced SOI provides the possibility to create and control quantum qubits, which is the main matter of concern in quantum computing.

%%%%%%%%%%%%%%%%%%%%%%%%%%%%%%%%%%%%%%%%%%%%%%%%%%%%%%%%%%%
%%%%%%%%%%%%%%%%%%%%%%%%%%%%%%%%%%%%%%%%%%%%%%%%%%%%%%%%%%%

We have considered a rotating electric field generated by applying time-dependent voltages to two sets of gates (configured
as sketched in Fig.~(\ref{sys})). Two comments are in order. The first is that a time-dependent electric field gives rise to a 
magnetic field according to the Maxwell equation
\begin{align}
\nabla \times {\bf B}(t) = \mu_0 {\bf J} + \frac{1}{c^2} \frac {\partial}{\partial t} {\bf E}(t).
\label{eqn1}
\end{align}
Here ${\bf B}$ and  ${\bf E}$ are the magnetic and electric field, respectively, ${\bf J}$ is the current density,
$\mu_0$ is the permeability of vacuum and $c$ is the speed of light.  

In estimating the strength of the 

generated {\bf B}-field, it is important to take the boundary conditions for Eq.~(\ref{eqn1}) into proper account.
We are not considering here an electromagnetic field traveling through free space but rather a situation 
where time-dependent gate voltages create the time-dependent, rotating (standing-wave) electric field given by Eq.~(\ref{n}).

Furthermore, we assume that we can neglect the first term on its right-hand side (the current density 
can be made arbitrarily small by decreasing the tunneling probability). We then find that the absolute value
of the right-hand side is
\begin{align}
\frac{\Omega}{c^2}E_0,
\end{align}
where $E_0$ is the amplitude of the rotating electric field.

To estimate the magnitude of the left-hand side of Eq.~(\ref{eqn1}), we note that the scale of the
spatial variation of the magnetic field is set by the distance between the gates, $L_g$, which in the device
used in Ref. \onlinecite{Scherubl2016} is given as 220 nm.
% = 2.2\, 10$^{-7}\,$m. 
Hence, for a rough estimate, we can say that
\begin{align}
\vert \nabla \times {\bf B} \vert \sim \eta \vert {\bf B} \vert /L_g,
\end{align}
where $\eta$  is a constant that depends on the precise geometry of the set-up. Consequently,
\begin{align}
\vert{\bf B}\vert \sim \frac{1}{\eta} \frac{L_g \Omega}{c^2}E_0 .
\label{ratio1}
\end{align}

In Ref.~\onlinecite{Scherubl2016} the side-gate-induced electric field inside the InAs nanowire is found to be 
$20\, {\rm mV/nm}$, 
%= 2\cdot 10^7\,{\rm V/m}$,
which we use 
% where V/m = N/C.  
%We will  use this value
 as an estimate of $E_0$. 
Furthermore we use $L_g \sim 0.2\,\mu{\rm m}$ and take $\eta = 1$ to get
\begin{align}
 \vert{\bf B}\vert \sim 
 %\left( \frac{\Omega}{2\pi \times \, 1\,{\rm THz}} \right) \frac{2\cdot 10^{-7}\cdot 2\pi \cdot 
% 10^{12}}{(3\cdot 10^8)^2}2\cdot 10^7  \sim 
\left( \frac{\Omega}{2\pi \times 1\, {\rm THz}} \right) 0.3\,{\rm mT}.
\label{Bestimate}
\end{align}
%so that the Zeeman splitting is
%\begin{align}
%\mu_B\vert{\bf B}\vert 
%%\sim \left( \frac{\Omega}{2\pi \times \, 1\,{\rm THz}} \right) 3\cdot 10^{-27}\,{\rm J}
%\sim  \left( \frac{\Omega}{2\pi \times \, 1\,{\rm THz}} \right) 0.02 \,\mu{\rm eV},
%\end{align}
%where $\mu_B=e\hbar/(2m_e) \sim 58 \times \mu$eV/T  is the Bohr magneton.
%% and $1\, {\rm eV} = 1.6 \cdot 10^{-19}\,{\rm J}$
%
%For the generated magnetic field to be negligible, this energy should be smaller than the energy 
%$k_B T$ at the temperature $T$ at which the experiment is
%done. Here the Boltzmann constant $k_B \sim 8.6 \cdot 10^{-5}\,{\rm eV/K}$, so that
%\begin{align}
%k_B T \sim 
%\left(\frac{T}{1\,{\rm mK}}\right) 10^{-26}\,{\rm J} 
%\left(\frac{T}{1\,{\rm mK}}\right) 
%1 \cdot 10^{-7} \,{\rm eV}.
%\end{align}
%
%We conclude that even at the very low temperature of 1 mK the {\bf B}-field can be neglected if 
%$\Omega < 2\pi \times 1\,{\rm THz}$ (since then $\mu_B\vert{\bf B}\vert \ll k_BT$).
%
We note that for any reasonable angular rotation frequency $\Omega$ 
this value is much smaller than what obtains for an electromagnetic wave traveling through
free space, in which case $\vert {\bf B}\vert = E_0/c \sim 0.1\,$T for the same electric field strength.

Both the magnetic field ${\bf B}$ given by Eq.~(\ref{Bestimate}) and the effective magnetic field 
${\bf B}_{\rm so}$ induced by the spin-orbit interaction affects the time evolution of any wave function
and their magnitudes should be compared to determine their relative importance. 
For real materials Eq.~(\ref{eq2}) for the latter,
%for ${\bf B}_{\rm so}$, 
although valid in free space, it is not really relevant. 
Instead, one typically considers the Rashba Hamiltonian, which, for our geometry, can be written as
\begin{align}
{\cal H}_{\rm so}  = \hbar k_{\rm so} v_F \sig \cdot (\hat {\bf x} \times \hat {\bf E}),
%= \alpha_R k_F \sig \cdot (\hat x \times \hat E),
\end{align}
where $v_F$ is the
%$\alpha_R = \hbar^2 k_{\rm so}/m^*$ is the Rashba coefficient and $k_F = m^* v_F/\hbar$ is the 
magnitude of the fermi velocity of the electrons 
%wave vector, while $m^*$ is the effective electron mass. 
By measuring, for example, the spin-orbit induced band splitting $\Delta_{\rm so} = 2 \hbar k_{\rm so} v_F$ 
one can then determine $k_{\rm so}$
%the Rashba coefficient 
and calculate the magnitude of the corresponding effective 
magnetic field as
\begin{align}
\vert {\bf B}_{\rm so} \vert %= \frac{\alpha_R k_F}{(g/2)\mu_B} 
= \frac{\hbar k_{\rm so} v_F}{(\vert g^* \vert /2)\mu_B}\ .
\end{align} 
Here $\mu_B$
%=e\hbar/(2m_e)$ 
%&\sim 9.3 \times 10^{-24} $J/T  
is the Bohr magneton and $g^*$ is the effective $g$-factor.
%$\hbar \sim 1.1 \times 10^{-34}\,$ Js is the reduce planck constant.
Using experimental numbers from Ref.~\onlinecite{Scherubl2016},
$k_{\rm so} = 1/(100\,{\rm nm})$ and $v_F \sim 4 \times 10^6\,{\rm m/s}$
\cite{vFcomment}, we find
\begin{align}
\vert {\bf B}_{\rm so} \vert \sim \frac{2}{\vert g^* \vert}\times 500\, {\rm T}\ .
\end{align} 
The value of the effective $g$-factor for the InAs nanowire studied was not given in 
Ref.~\onlinecite{Scherubl2016}, but other works on InAs nanowire quantum dots \cite{Fasth2007} indicate it might be 
of order 10 ($\vert g^* \vert \sim 10$). Nevertheless, for any realistic value $ \vert {\bf B}_{\rm so} \vert \gg \vert {\bf B} \vert$,
so the magnetic field induced by the time-dependence of the electric field can be neglected.

The second comment we want to make is that, as an alternative to creating a time-dependent electric
field by applying time-dependent gate voltages as described above, one may irradiate one or more gates
with microwaves that arrive at the gates via some waveguide(s) or stripline(s). 
It is important that the gates and the waveguides/striplines prevent the microwave field from 
directly reaching the active area containing the quantum dot and the weak links. Then, if
the electrodes are thick enough and the microwave frequency is lower than the plasma frequency of the gate metal, the skin effect will confine the magnetic component of the microwave as well as the 
electric component that is parallel to the metal gate to the gate surface, so that the active part of the
device is shielded. The electric component of the microwave field that is perpendicular to the gate
surface will be screened by charges on both sides of the gate. The time-dependent charge distribution
on the back side of the gate 
will then create a time-dependent electric field of about the same magnitude as the perpendicular 
electric component of the impinging microwave. This electric field will affect the weak link in much the same way as 
in the case of voltage-driven gates.

%%%%%%%%%%%%%%%%%%%%%%%%%%%%%%%%%%%%%%%%%%%%%%%%%%%%%%%%%%%
%%%%%%%%%%%%%%%%%%%%%%%%%%%%%%%%%%%%%%%%%%%%%%%%%%%%%%%%%%%

In conclusion, we show that a time-periodic rotating electric field, acting on a weak link between two reservoirs, generates a nondissipative magnetization along the weak link direction. This magnetization is not present in the other two directions; in those, the magnetization is purely time-driven. The time-independent magnetization we achieve is estimated in a scenario where the Fermi functions of the two reservoirs are equal or differ by a bias voltage, providing estimates for its magnitude in an open circuit.  

Igor Kulik is a prominent name in solid-state physics. His contribution to the mesoscopic physics of normal and superconducting conductors continues to be the focus of modern research. One of us (R.S.) was fortunate to begin his scientific career under Kulik's supervision and benefited from scientific collaboration with him, working on the Coulomb blockade of single-electron tunneling and the theory of point contact spectroscopy in metals. O.E.W. collaborated with him in 1981 during a stay at UCLA (as a postdoc with R. L. Orbach). Although the interaction was short, it produced a paper entitled {\it Pair susceptibility and mode propagation in superconductors: a microscopic approach} which is still cited. %J. Low Temp. Phys. 43, 591-620 (1981)
 A very good memory of those days makes us feel very honored and happy to contribute to this special issue of Fiz. Nizk. Temp.
 
% {\color{blue} [MJ: In the last sentence above "special" is used twice and we refer to LTP, which is the name of the translated journal but not the journal we will submit to. How about:
%  "... to contribute to this special issue of Fiz. Nizk. Temp."?]}
 
%%%%%%%%%%%%%%%%%%%%%%%%%%%%%%%%%%%%%%%%%%%%%%%%%%%%%%%%%%%
%%%%%%%%%%%%%%%%%%%%%%%%%%%%%%%%%%%%%%%%%%%%%%%%%%%%%%%%%%%

\begin{acknowledgments}
OEW, AA, RIS, and MJ acknowledge the hospitality of the PCS at IBS, Daejeon, S. Korea, where part of this work was  supported by
IBS funding number (IBS-R024-D1). DC acknowledges the financial support from DST with project number: DST/WISE-PDF/PM-40/2023.
 \end{acknowledgments}

\onecolumngrid

\appendix

%%%%%%%%%%%%%%%%%%%%%%%%%%%%%%%%%%%%%%%%%%%%%%%%%
%%%%%%%%%%%%%%%%%%%%%%%%%%%%%%%%%%%%%%%%%%%%%%%%%
\section{Technical Details}
\label{App}

%%%%%%%%%%%%%%%%%%%%%%%%%%%%%%%%%%%%%%%%%%%%%%%%%%%%%%%%%%%
%%%%%%%%%%%%%%%%%%%%%%%%%%%%%%%%%%%%%%%%%%%%%%%%%%%%%%%%%%%

\subsection{ The Green's functions} 

\label{GF}

The Keldysh technique requires calculations of the retarded, advanced, and lesser Green's functions.
The three Keldysh functions on the dot, $G^{r,a,<}_{dd}(t,t')$, are obtained from the Dyson equation 
 \begin{align}        
G^{}_{dd}(t,t')=g^{}_{d}(t,t')+\int\int dt^{}_{1}dt^{}_{2}
g^{}_{d}(t,t^{}_{1})\Sigma^{}_{}(t^{}_{1},t^{}_{2})G^{}_{dd}(t^{}_{2},t')\ ,
\label{DyGdd}
\end{align} 
where $g_{d}$ is the Green's function of the decoupled dot, and the self-energy $\Sigma$ is given in Eqs. (\ref{sigral}). We assume that $\epsilon_{d}$ is larger than the mean chemical potential of the two reservoirs, rendering the decoupled dot empty of electrons.
By Langreth's rules \cite{Haug2008, Langreth1976}, Eq.~(\ref{DyGdd}) yields [see, for instance, Ref.~\cite{OEW2017})
 \begin{align}
G^{r,a}_{dd}(t,t')=\mp i\Theta (\pm t\mp t')e^{-i(\epsilon^{}_{d}\mp i\Gamma)(t-t')}\ , 
\label{Gra}
\end{align}
and  
\begin{align}
&G^{<}_{dd}(t^{}_{1},t^{}_{2})=\int dt'^{}_{1}\int dt'^{}_{2}G^{r}_{dd}(t^{}_{1},t'^{}_{1})\Sigma^{<}(t'^{}_{1},t'^{}_{2})G^{a}_{dd}(t'^{}_{2},t^{}_{2})\nonumber\\
&=i \sum_{\bf k}f^{}_{L}(\epsilon^{}_{k})e^{-i\epsilon^{}_{k}(t^{}_{1}-t^{}_{2})}\int^{t^{}_{1}}dt'^{}_{1}\oJ^{\dagger}_{L}(t'^{}_{1})e^{i(\epsilon^{}_{d}-\epsilon^{}_{k}-i\Gamma)(t'^{}_{1}-t^{}_{1})}\int^{t^{}_{2}}dt'^{}_{2}\oJ^{}_{L}(t'^{}_{2})e^{-i(\epsilon^{}_{d}-\epsilon^{}_{k}+i\Gamma)(t'^{}_{2}-t^{}_{2})}+[L\rightarrow R]\ ,
\end{align}
where in the second equality, we have used Eqs. (\ref{sigral}) and (\ref{Gra}). Introducing the notations
\begin{align}
\oJ^{}_{L,-}(\epsilon^{}_{k},t)\equiv\int^{t} dt'e^{i(\epsilon^{}_{d}-\epsilon^{}_{k}-i\Gamma)(t'-t)}\oJ^{\dagger}_{L}(t')&=-iJ^{\dagger}_{L,0}\Big (\frac{C^{}_{L}}{\epsilon^{}_{d}-\epsilon^{}_{k}-i\Gamma}\nonumber\\
&-iS^{}_{L}\Big [\frac{\sigma^{}_{-}e^{i\Omega t}}{\epsilon^{}_{d}-\epsilon^{}_{k}+\Omega-i\Gamma}+\frac{\sigma^{}_{+}e^{-i\Omega t}}{\epsilon^{}_{d}-\epsilon^{}_{k}-\Omega-i\Gamma}\Big]\Big )\ ,\nonumber\\
\oJ^{}_{R,-}(\epsilon^{}_{p},t)\equiv\int^{t} dt'e^{i(\epsilon^{}_{d}-\epsilon^{}_{p}-i\Gamma)(t'-t)}\oJ^{\dagger}_{R}(t')&=-iJ^{\dagger}_{R,0}\Big (\frac{C^{}_{R}}{\epsilon^{}_{d}-\epsilon^{}_{p}-i\Gamma}\nonumber\\
&+iS^{}_{R}\Big [\frac{\sigma^{}_{-}e^{i\Omega t}}{\epsilon^{}_{d}-\epsilon^{}_{p}+\Omega-i\Gamma}+\frac{\sigma^{}_{+}e^{-i\Omega t}}{\epsilon^{}_{d}-\epsilon^{}_{p}-\Omega-i\Gamma}\Big]\Big )\ ,
\label{JLRt}
\end{align} 
with $\oJ^{}_{L(R),+}=[\oJ^{}_{L(R),-}]^{\dagger}$, and using for brevity
\begin{align}
C^{}_{L(R)}\equiv \cos(k^{}_{\rm so}d^{}_{L(R)})\ ,\ \ \ 
S^{}_{L(R)}\equiv \sin(k^{}_{\rm so}d^{}_{L(R)}) \ ,
\label{CS}
\end{align}
one finds
\begin{align}
G^{<}_{dd}(t^{}_{1},t^{}_{2})
=& i \sum_{\bf k}f^{}_{L}(\epsilon^{}_{k})e^{-i\epsilon^{}_{k}(t^{}_{1}-t^{}_{2})}\oJ^{}_{L,-}(\epsilon^{}_{k},t^{}_{1})\oJ^{}_{L,+}(\epsilon^{}_{k},t^{}_{2})\nonumber\\
&+
i \sum_{\bf p}f^{}_{R}(\epsilon^{}_{p})e^{-i\epsilon^{}_{p}(t^{}_{1}-t^{}_{2})}\oJ^{}_{R,-}(\epsilon^{}_{p},t^{}_{1})\oJ^{}_{R,+}(\epsilon^{}_{p},t^{}_{2})
\ .
\label{Gdo}
\end{align} 
Note that $\oJ^{}_{L(R),\pm}$ depends on $\epsilon^{}_{k}(\epsilon^{}_{p})$, as opposed to $\oJ^{}_{L(R)}(t)$,  and is dimensionless. 

%%%%%%%%%%%%%%%%%%%%%%%%%%%%%%%%%%%%%%%%%%%%%%%%%%%%%%%%%%%

The equal-time Green's function $G^{<}_{dd}(t,t)$ is needed for deriving the magnetizations generated on various parts of the device. Its explicit form is presented for slow AC variation, $\Omega<<\Gamma$, up to linear order in $\Omega$, 
\begin{align}
G^{<}_{dd}(t,t)
&\approx
 i\frac{\Gamma^{}_{L}}{\pi}\int d\epsilon^{}_{k}f^{}_{L}(\epsilon^{}_{k})\Big (\frac{1}{(\epsilon^{}_{d}-\epsilon^{}_{k})^{2}+\Gamma^{2}}+\frac{2\Omega(\epsilon^{}_{d}-\epsilon^{}_{k})S^{}_{L}}{[(\epsilon^{}_{d}-\epsilon^{}_{k})^{2}+\Gamma^{2}]^{2}}[S^{}_{L}\hat{\bf x}-C^{}_{L}\hat{\bf n}(t)]\cdot\sig\Big)\nonumber\\
 &+i\frac{\Gamma^{}_{R}}{\pi}\int d\epsilon^{}_{p}f^{}_{R}(\epsilon^{}_{p})\Big (\frac{1}{(\epsilon^{}_{d}-\epsilon^{}_{p})^{2}+\Gamma^{2}}+\frac{2\Omega(\epsilon^{}_{d}-\epsilon^{}_{p})S^{}_{R}}{[(\epsilon^{}_{d}-\epsilon^{}_{p})^{2}+\Gamma^{2}]^{2}}[S^{}_{R}\hat{\bf x}+C^{}_{R}\hat{\bf n}(t)]\cdot\sig\Big) \ .
 \label{Gdtt}
  %\Big [S^{}_{L}\sigma^{}_{x}-C^{}_{L}[\sigma^{}_{y}\cos(\Omega t)-\sigma^{}_{z}\sin(\Omega t)]\Big ]\Big)+\{L\rightarrow R\}\ ,
\end{align}
%where we have introduced the unit vectors
%\begin{align}
%\hat{\bf u}^{}_{L}(t)\cdot\sig&=S^{}_{L}\sigma^{}_{x}-C^{}_{L}[\hat{\bf y}\sin(\Omega t)-\hat{\bf z}\cos(\Omega t)]\cdot\sig\ ,\nonumber\\
%\hat{\bf u}^{}_{R}(t)\cdot\sig&=S^{}_{R}\sigma^{}_{x}+C^{}_{R}[\hat{\bf y}\sin(\Omega t)-\hat{\bf z}\cos(\Omega t)]\cdot\sig\ ,
%\end{align}
Apart from a time-independent component along the weak link direction $\hat{\bf x}$, the other components of the unit vectors multiplying $\sig$ are parallel to that describing the circularly polarized AC field,  
\begin{align}
\hat{\bf n}(t)=\hat{\bf y}\sin(\Omega t)-\hat{\bf z}\cos(\Omega t)\ .
\label{cpn}
\end{align}
It is interesting to compare the structure of the two terms in the circular brackets in Eq.~(\ref{Gdtt}). The first ones correspond to the Breit-Wigner 
resonances, $[(\epsilon^{}_{d}-\epsilon^{}_{k(p)})^{2}+\Gamma^{2}]^{-1}$ formed on the dot. As seen, for low enough AC frequency, $\Omega$, these are not affected by the spin-orbit coupling. On the other hand, the second terms that vanish for zero SOI correspond to the derivatives of the Breit-Wigner resonance,  $2(\epsilon^{}_{d}-\epsilon^{}_{k(p)})[(\epsilon^{}_{d}-\epsilon^{}_{k(p)})^{2}+\Gamma^{2}]^{-2}=(d/d\epsilon^{}_{k(p)})[(\epsilon^{}_{d}-\epsilon^{}_{k(p)})^{2}+\Gamma^{2}]^{-1}$.

%%%%%%%%%%%%%%%%%%%%%%%%%%%%%%%%%%%%%%%%%%%%%%%%%%%%%%%%%%%
%%%%%%%%%%%%%%%%%%%%%%%%%%%%%%%%%%%%%%%%%%%%%%%%%%%%%%%%%%%

The Dyson equation for the Green's function of the left reservoir, $G_{{\bf k}{\bf k'}}$ (a matrix in spin space), reads
\begin{align}
\sum_{{\bf k},{\bf k}'}G^{}_{{\bf k}{\bf k}'}(t,t)
=\sum_{{\bf k}}g^{}_{{\bf k}}(t,t)\delta^{}_{{\bf k},{\bf k}'}&+\sum_{{\bf k},{\bf k}'}\int dt^{}_{1}\int dt^{}_{2}g^{}_{\bf k}(t,t^{}_{1})
\oJ^{}_{L}(t^{}_{1})G^{}_{dd}(t^{}_{1},t^{}_{2})\oJ^{\dagger}_{L}(t^{}_{2})g^{}_{{\bf k}'}(t^{}_{2},t)\ .
\label{sGkkp}
\end{align}
Applying Langreth's rules \cite{Haug2008, Langreth1976} to Eq.~(\ref{sGkkp}),
one finds, 
\begin{align}
\sum_{{\bf k},{\bf k}'}G^{<}_{{\bf k}{\bf k}'}(t,t)
=\sum_{{\bf k}}g^{<}_{{\bf k}}(t,t)\delta^{}_{{\bf k},{\bf k}'}&+\sum_{{\bf k},{\bf k}'}\int dt^{}_{1}\int dt^{}_{2}g^{<}_{\bf k}(t,t^{}_{1})
\oJ^{}_{L}(t^{}_{1})G^{a}_{dd}(t^{}_{1},t^{}_{2})\oJ^{\dagger}_{L}(t^{}_{2})g^{a}_{{\bf k}'}(t^{}_{2},t)
\nonumber\\
&+\sum_{{\bf k},{\bf k}'}\int dt^{}_{1}\int dt^{}_{2}g^{r}_{{\bf k}}(t,t^{}_{1})
\oJ^{}_{L}(t^{}_{1})G^{r}_{dd}(t^{}_{1},t^{}_{2})\oJ^{\dagger}_{L}(t^{}_{2})g^{<}_{{\bf k}'}(t^{}_{2},t)\nonumber\\
&+\sum_{{\bf k},{\bf k}'}\int dt^{}_{1}\int dt^{}_{2}g^{r}_{{\bf k}}(t,t^{}_{1})
\oJ^{}_{L}(t^{}_{1})G^{<}_{dd}(t^{}_{1},t^{}_{2})\oJ^{\dagger}_{L}(t^{}_{2})g^{a}_{{\bf k}'}(t^{}_{2},t)\ , 
\label{LLM}
\end{align}
with an analogous expression for $\sum_{{\bf p},{\bf p}'}G^{<}_{{\bf p}{\bf p}'}(t,t)$. 
The retarded, advanced, and lesser Green's functions of the decoupled (left) lead are functions of $\epsilon^{}_{k}$, 
\begin{align}
g^{r,a}_{k}(t,t')&=\mp i\Theta (\pm t\mp t')e^{-i\epsilon^{}_{k}(t-t')}\ ,\ \ 
g^{<}_{k}(t,t')=if^{}_{L}(\epsilon^{}_{k})
e^{-i\epsilon^{}_{k}(t-t')}=f^{}_{L}(\epsilon^{}_{k})[g^{a}_{k}(t,t')-g^{r}_{k}(t,t')]\ .
\label{gk}
\end{align}
Therefore, the second term on the right-hand-side of Eq.~(\ref{LLM}) necessitates
$t>t^{}_{2}>t^{}_{1}$  (and not only $t^{}_{1}<t$)
and the third  term there requires
$t>t^{}_{1}>t^{}_{2}$ (and not only $t^{}_{2}<t$). On the other hand, the fourth term in Eq.~(\ref{LLM}) requires just $t>t^{}_{1}$ and $t>t^{}_{2}$.
This means that the ``area"  covered by the double integration in the fourth term equals the sum of the ``areas" covered in the second and third terms. 
By exploiting Eqs. (\ref{gk}), one finds
\begin{align}
&\frac{d}{dt}g^{<}_{{\bf k}}(t,t^{}_{1})g^{a}_{{\bf k}'}(t^{}_{2},t)=i(\epsilon^{}_{k'}-\epsilon^{}_{k})g^{<}_{{\bf k}}(t,t^{}_{1})g^{a}_{{\bf k}'}(t^{}_{2},t)+i\delta(t-t^{}_{2})g^{<}_{{\bf k}}(t,t^{}_{1})\ ,\nonumber\\
&\frac{d}{dt}g^{<}_{{\bf k}'}(t^{}_{2},t)g^{r}_{{\bf k}}(t,t^{}_{1})=i(\epsilon^{}_{k'}-\epsilon^{}_{k})g^{<}_{{\bf k}'}(t^{}_{2},t)g^{r}_{{\bf k}}(t,t^{}_{1})-i\delta(t-t^{}_{1})g^{<}_{{\bf k}}(t^{}_{2},t)\ ,\nonumber\\&\frac{d}{dt}g^{r}_{{\bf k}}(t,t^{}_{1})g^{a}_{{\bf k}'}(t^{}_{2},t)=i(\epsilon^{}_{k'}-\epsilon^{}_{k})g^{r}_{{\bf k}}(t,t^{}_{1})g^{a}_{{\bf k}'}(t^{}_{2},t)-i\delta(t-t^{}_{1})g^{a}_{{\bf k}'}(t^{}_{2},t)+i\delta(t-t^{}_{2})g^{r}_{{\bf k}'}(t,t^{}_{1})\ ,
\end{align}
from which follows Eq.~(\ref{derGkk}) in Sec. \ref{prob}. 
In particular, 
$d\sum_{{\bf k}}G^{<}_{{\bf k}{\bf k}}(t,t)/dt$, the cornerstone for deriving the charge current and {\it part} of the magnetization fluxes (see the discussion in Sec. \ref{prob}), is given by
\begin{align}
\frac{d}{dt}\sum_{{\bf k}}G^{<}_{{\bf k}{\bf k}}(t,t)&=i\sum_{{\bf k}}\int dt^{}_{1}[g^{r}_{{\bf k}}(t,t^{}_{1})
\oJ^{}_{L}(t^{}_{1})G^{<}_{dd}(t^{}_{1},t)\oJ^{\dagger}_{L}(t)+g^{<}_{\bf k}(t,t^{}_{1})
\oJ^{}_{L}(t^{}_{1})G^{a}_{dd}(t^{}_{1},t)\oJ^{\dagger}_{L}(t)]\nonumber\\
&-i\sum_{{\bf k}}\int dt^{}_{1}[\oJ^{}_{L}(t)G^{<}_{dd}(t,t^{}_{1})\oJ^{\dagger}_{L}(t^{}_{1})g^{a}_{{\bf k}}(t^{}_{{1}},t)+\oJ^{}_{L}(t)G^{r}_{dd}(t,t^{}_{1})\oJ^{\dagger}_{L}(t^{}_{1})g^{<}_{{\bf k}}(t^{}_{{1}},t)]\ .
\label{dGkk}
\end{align}
It is interesting to observe that in Eq.~(\ref{dGkk}), we don't have any contribution from the first term of Eq.~(\ref{LLM}). This is due to the fact that for ${\rm k}={\rm k'}$, the first term, i.e, the time derivative of $g^{<}_{{\rm k}}(t,t)$ is zero as $g^{<}_{{\rm k}}(t,t)$ is time independent.
Within the wide-band  limit, this expression becomes [see Eq.~(\ref{del})]
\begin{align}
\frac{d}{dt}\sum_{{\bf k}}G^{<}_{{\bf k}{\bf k}}(t,t)&\approx-i\sum_{{\bf k}}f^{}_{L}(\epsilon^{}_{k})[\oJ^{}_{L,+}(\epsilon^{}_{k},t)\oJ^{\dagger}_{L}(t)+
\oJ^{}_{L}(t)\oJ^{\dagger}_{L,+}(\epsilon^{}_{k},t)] +2\pi{\cal N}^{}_{L}\oJ^{}_{L}(t)G^{<}_{dd}(t,t)\oJ^{\dagger}_{L}(t)\ ,
\label{Gkkdt}
\end{align}
where we have used Eq.~(\ref{Gra}). In the $\Omega\ll \Gamma$ limit
\begin{align}
\frac{d}{dt}\sum_{{\bf k}}G^{<}_{{\bf k}{\bf k}}(t,t)&\approx-i\frac{2\Gamma^{}_{L}\Gamma^{}_{R}}{\pi}\Big\{\int d\epsilon^{}_{k}f^{}_{L}(\epsilon^{}_{k})\Big (\frac{1}{(\epsilon^{}_{d}-\epsilon^{}_{k})^{2}+\Gamma^{2}}
-\frac{2\Omega(\epsilon^{}_{d}-\epsilon^{}_{k})S^{}_{L}}{[(\epsilon^{}_{d}-\epsilon^{}_{k})^{2}+\Gamma^{2}]^{2}}[S^{}_{L}\hat{\bf x}+C^{}_{L}\hat{\bf n}(t)]\cdot\sig\Big)
\nonumber\\
%&+2\Gamma^{}_{L}\Big \{i\frac{\Gamma^{}_{L}}{\pi}\int d\epsilon^{}_{k}f^{}_{L}(\epsilon^{}_{k})\Big(\frac{1}{(\epsilon^{}_{d}-\epsilon^{}_{k})^{2}+\Gamma^{2}}-\frac{2\Omega(\epsilon^{}_{d}-\epsilon^{}_{k})S^{}_{L}}{[(\epsilon^{}_{d}-\epsilon^{}_{k})^{2}+\Gamma^{2}]^{2}}[S^{}_{L}\hat{\bf x}+C^{}_{L}\hat{\bf n}(t)]\cdot\sig\Big)\nonumber\\
%&+i\frac{\Gamma^{}_{R}}{\pi}
&-\int d\epsilon^{}_{p}f^{}_{R}(\epsilon^{}_{p})\Big(\frac{1}{(\epsilon^{}_{d}-\epsilon^{}_{p})^{2}+\Gamma^{2}}+\frac{2\Omega(\epsilon^{}_{d}-\epsilon^{}_{p})S^{}_{R}}{[(\epsilon^{}_{d}-\epsilon^{}_{p})^{2}+\Gamma^{2}]^{2}}[S^{}_{2L+R}\hat{\bf x}+C^{}_{2L+R}\hat{\bf n}(t)]\cdot\sig\Big)\Big\}\ ,
\label{derGL}
\end{align}
%\begin{align}
%\frac{d}{dt}\sum_{{\bf k}}G^{<}_{{\bf k}{\bf k}}(t,t)&\approx-i\frac{2\Gamma^{}_{L}\Gamma}{\pi}\int d\epsilon^{}_{k}f^{}_{L}(\epsilon^{}_{k})\Big (\frac{1}{(\epsilon^{}_{d}-\epsilon^{}_{k})^{2}+\Gamma^{2}}
%-\frac{2\Omega(\epsilon^{}_{d}-\epsilon^{}_{k})S^{}_{L}}{[(\epsilon^{}_{d}-\epsilon^{}_{k})^{2}+\Gamma^{2}]^{2}}[S^{}_{L}\hat{\bf x}+C^{}_{L}\hat{\bf n}(t)]\cdot\sig\Big)\ .
%\nonumber\\
%&+2\Gamma^{}_{L}\Big \{i\frac{\Gamma^{}_{L}}{\pi}\int d\epsilon^{}_{k}f^{}_{L}(\epsilon^{}_{k})\Big(\frac{1}{(\epsilon^{}_{d}-\epsilon^{}_{k})^{2}+\Gamma^{2}}-\frac{2\Omega(\epsilon^{}_{d}-\epsilon^{}_{k})S^{}_{L}}{[(\epsilon^{}_{d}-\epsilon^{}_{k})^{2}+\Gamma^{2}]^{2}}[S^{}_{L}\hat{\bf x}+C^{}_{L}\hat{\bf n}(t)]\cdot\sig\Big)\nonumber\\
%&+i\frac{\Gamma^{}_{R}}{\pi}\int d\epsilon^{}_{p}f^{}_{R}(\epsilon^{}_{p})\Big(\frac{1}{(\epsilon^{}_{d}-\epsilon^{}_{p})^{2}+\Gamma^{2}}+\frac{2\Omega(\epsilon^{}_{d}-\epsilon^{}_{p})S^{}_{R}}{[(\epsilon^{}_{d}-\epsilon^{}_{p})^{2}+\Gamma^{2}]^{2}}[S^{}_{2L+R}\hat{\bf x}+C^{}_{2L+R}\hat{\bf n}(t)]\cdot\sig\Big)\Big\}\ ,
%\label{derGL}
%\end{align}
where
\begin{align}
S^{}_{2L+R}=\sin(k^{}_{\rm so}[2d^{}_{L}+d^{}_{R}])\ ,\ \ C^{}_{2L+R}=\cos(k^{}_{\rm so}[2d^{}_{L}+d^{}_{R}])\ . 
\label{2L+R}
\end{align}
Similarly, 
\begin{align}
\frac{d}{dt}\sum_{{\bf p}}G^{<}_{{\bf p}{\bf p}}(t,t)&\approx-i\frac{2\Gamma^{}_{R}\Gamma^{}_{L}}{\pi}\Big\{\int d\epsilon^{}_{p}f^{}_{R}(\epsilon^{}_{p})\Big (\frac{1}{(\epsilon^{}_{d}-\epsilon^{}_{p})^{2}+\Gamma^{2}}
-\frac{2\Omega(\epsilon^{}_{d}-\epsilon^{}_{p})S^{}_{R}}{[(\epsilon^{}_{d}-\epsilon^{}_{p})^{2}+\Gamma^{2}]^{2}}[S^{}_{R}\hat{\bf x}-C^{}_{R}\hat{\bf n}(t)]\cdot\sig\Big)
\nonumber\\
%&+2\Gamma^{}_{R}\Big \{\frac{i\Gamma^{}_{L}}{\pi}
&-\int d\epsilon^{}_{k}f^{}_{L}(\epsilon^{}_{k})\Big (\frac{1}{(\epsilon^{}_{d}-\epsilon^{}_{k})^{2}+\Gamma^{2}}+\frac{2\Omega(\epsilon^{}_{d}-\epsilon^{}_{k})S^{}_{L}}{[(\epsilon^{}_{d}-\epsilon^{}_{k})^{2}+\Gamma^{2}]^{2}}[S^{}_{2R+L}\hat{\bf x}%\sin(k^{}_{\rm so}[2d^{}_{R}+d^{}_{L}])-\cos(k^{}_{\rm so}[2d^{}_{R}+d^{}_{L}])
-C^{}_{2R+L}\hat{\bf n}(t)]\cdot\sig\Big )\Big\}\ ,%\nonumber\\
%&+i\frac{\Gamma^{}_{R}}{\pi}\int d\epsilon^{}_{p}f^{}_{R}(\epsilon^{}_{p})\Big (\frac{1}{(\epsilon^{}_{d}-\epsilon^{}_{p})^{2}+\Gamma^{2}}-\frac{2\Omega(\epsilon^{}_{d}-\epsilon^{}_{p})S^{}_{R}}{[(\epsilon^{}_{d}-\epsilon^{}_{p})^{2}+\Gamma^{2}]^{2}}[S^{}_{R}\hat{\bf x}%\sin(k^{}_{\rm so}d^{}_{R})-
%\cos(k^{}_{\rm so}d^{}_{R})
%-C^{}_{R}\hat{\bf n}(t)]\cdot\sig\Big )\Big\}\ ,
\label{derGR}
\end{align}
%\begin{align}
%\frac{d}{dt}\sum_{{\bf p}}G^{<}_{{\bf p}{\bf p}}(t,t)&\approx-i\frac{2\Gamma^{}_{R}\Gamma}{\pi}\int d\epsilon^{}_{p}f^{}_{R}%(\epsilon^{}_{p})\Big (\frac{1}{(\epsilon^{}_{d}-\epsilon^{}_{p})^{2}+\Gamma^{2}}
%-\frac{2\Omega(\epsilon^{}_{d}-\epsilon^{}_{p})S^{}_{R}}{[(\epsilon^{}_{d}-\epsilon^{}_{p})^{2}+\Gamma^{2}]^{2}}[S^{}_{R}\hat{\bf %x}-C^{}_{R}\hat{\bf n}(t)]\cdot\sig\Big)\ .
%\nonumber\\
%&+2\Gamma^{}_{R}\Big \{\frac{i\Gamma^{}_{L}}{\pi}\int d\epsilon^{}_{k}f^{}_{L}(\epsilon^{}_{k})\Big (\frac{1}{(\epsilon^{}_{d}-%\epsilon^{}_{k})^{2}+\Gamma^{2}}+\frac{2\Omega(\epsilon^{}_{d}-\epsilon^{}_{k})S^{}_{L}}{[(\epsilon^{}_{d}-\epsilon^{}_{k})^{2}+%\Gamma^{2}]^{2}}[S^{}_{2R+L}\hat{\bf x}%\sin(k^{}_{\rm so}[2d^{}_{R}+d^{}_{L}])-\cos(k^{}_{\rm so}[2d^{}_{R}+d^{}_{L}])
%-C^{}_{2R+L}\hat{\bf n}(t)]\cdot\sig\Big )\nonumber\\
%&+i\frac{\Gamma^{}_{R}}{\pi}\int d\epsilon^{}_{p}f^{}_{R}(\epsilon^{}_{p})\Big (\frac{1}{(\epsilon^{}_{d}-\epsilon^{}_{p})^{2}+%\Gamma^{2}}-\frac{2\Omega(\epsilon^{}_{d}-\epsilon^{}_{p})S^{}_{R}}{[(\epsilon^{}_{d}-\epsilon^{}_{p})^{2}+\Gamma^{2}]^{2}}%[S^{}_{R}\hat{\bf x}%\sin(k^{}_{\rm so}d^{}_{R})-
%\cos(k^{}_{\rm so}d^{}_{R})
%-C^{}_{R}\hat{\bf n}(t)]\cdot\sig\Big )\Big\}\ ,
%\label{derGR}
%\end{align}
with
\begin{align}
S^{}_{2R+L}=\sin(k^{}_{\rm so}[2d^{}_{R}+d^{}_{L}])\ ,\ \ C^{}_{2R+L}=\cos(k^{}_{\rm so}[2d^{}_{R}+d^{}_{L}])\ . 
\label{2R+L}
\end{align}

%%%%%%%%%%%%%%%%%%%%%%%%%%%%%%%%%%%%%%%%%%%%%%%%%%%%%%%%%%%
%%%%%%%%%%%%%%%%%%%%%%%%%%%%%%%%%%%%%%%%%%%%%%%%%%%%%%%%%%%

The processes that do not conserve momentum result, within the wide-band limit, in the expression 
\begin{align}
\sum_{{\bf k}\neq{\bf k}'}G^{<}_{{\bf k}{\bf k}'}(t,t)&\approx
-i\pi{\cal N}^{}_{L}\sum_{\bf k}f^{}_{L}(\epsilon^{}_{k})\Big[\oJ^{}_{L,+}(\epsilon^{}_{k},t)
\oJ^{\dagger}_{L}(t)+\oJ^{}_{L}(t)\oJ^{}_{L,-}(\epsilon^{}_{k},t)\Big]\nonumber\\
&+2i(\pi{\cal N}^{}_{L})^{2}\sum_{{\bf k}}f^{}_{L}(\epsilon^{}_{k})\oJ^{}_{L}(t)\oJ^{}_{L,-}(\epsilon^{}_{k},t)\oJ^{}_{L,+}(\epsilon^{}_{k},t)\oJ^{\dagger}_{L}(t^{}_{})\nonumber\\
&+2i(\pi{\cal N}^{}_{L})^{2}\sum_{{\bf p}}f^{}_{R}(\epsilon^{}_{p})\oJ^{}_{L}(t)\oJ^{}_{R,-}(\epsilon^{}_{p},t)\oJ^{}_{R,+}(\epsilon^{}_{p},t)\oJ^{\dagger}_{L}(t^{}_{})\ .
\label{Gnc}
\end{align}
It is to be noted from Eq.~(\ref{LLM}) that the first term is not included in Eq.~(\ref{Gnc}). The reason is that in Eq.~(\ref{Gnc}), we have only considered the momentum non-conserving processes i.e where ${\rm k}\neq {\rm k}'.$ The first term in Eq.~(\ref{LLM}) is only non-zero for ${\rm k}={\rm k}'$.
It is interesting to compare this expression with the one presented in Eq.~(\ref{Gkkdt}) for $d\sum_{{\bf k}}G^{<}_{\bf kk}(t,t)/dt$. Inserting the explicit expression for $G^{<}_{dd}(t,t)$, Eq.~(\ref{Gdtt}), into Eq.~(\ref{Gkkdt}) one finds that within the wide-band limit, 
\begin{align}
\sum_{{\bf k}\neq{\bf k}'}G^{<}_{{\bf k}{\bf k}'}(t,t)=\pi{\cal N}^{}_{L}\frac{d}{dt}\sum_{{\bf k}}G^{<}_{{\bf k}{\bf k}}(t,t)\ .
\label{equ}
\end{align}
%%%%%%%%%%%%%%%%%%%%%%%%%%%%%%%%%%%%%%%%%%%%%%%%%%%%%%%%%%%
%%%%%%%%%%%%%%%%%%%%%%%%%%%%%%%%%%%%%%%%%%%%%%%%%%%%%%%%%%%
%%%%%%%%%%%%%%%%%%%%%%%%%%%%%%%%%%%%%%%%%%%%%%%%%%%%%%%%%%%

Another interesting aspect of the wide-band approximation emerges upon inspecting the Green's function $G_{{\bf k}{\bf k}'}$, Eq. (\ref{LLM}) (an analogous observation holds for $G_{{\bf p}{\bf p}'}$),  specifically the last three terms there, that result from the coupling with the dot. Whereas it is straightforward to apply this approximation to $G^{<}_{{\bf k}\neq{\bf k}'}$ as done above, applying it to $G^{<}_{{\bf k}{\bf k}}$ reveals a caveat. One notes that [see Eqs. (\ref{gk})]
\begin{align}
&\sum_{\bf k}g^{<}_{{\bf k}}(t,t^{}_{1})g^{a}_{\bf k}(t^{}_{2},t)
 =-\Theta(t-t^{}_{2})\sum_{\bf k}f^{}_{L}(\epsilon^{}_{k})e^{-i\epsilon^{}_{k}(t-t^{}_{1})}e^{-i\epsilon^{}_{k}(t^{}_{2}-t)}
 =-\Theta(t-t^{}_{2})\sum_{\bf k}f^{}_{L}(\epsilon^{}_{k})e^{-i\epsilon^{}_{k}(t^{}_{2}-t^{}_{1})}\ ,\nonumber\\
&\sum_{\bf k}g^{r}_{{\bf k}}(t,t^{}_{1})g^{<}_{\bf k}(t^{}_{2},t)
 =\Theta(t-t^{}_{1})\sum_{\bf k}f^{}_{L}(\epsilon^{}_{k})e^{-i\epsilon^{}_{k}(t-t^{}_{1})}e^{-i\epsilon^{}_{k}(t^{}_{2}-t)}
 =\Theta(t-t^{}_{1})\sum_{\bf k}f^{}_{L}(\epsilon^{}_{k})e^{-i\epsilon^{}_{k}(t^{}_{2}-t^{}_{1})}\ , \nonumber\\ 
 &\sum_{\bf k}g^{r}_{{\bf k}}(t,t^{}_{1})g^{a}_{\bf k}(t^{}_{2},t)
 =\Theta(t-t^{}_{1})\Theta(t-t^{}_{2})\sum_{\bf k}e^{-i\epsilon^{}_{k}(t-t^{}_{1})}e^{-i\epsilon^{}_{k}(t^{}_{2}-t)}
 =\Theta(t-t^{}_{1})\Theta(t-t^{}_{2})\sum_{\bf k}e^{-i\epsilon^{}_{k}(t^{}_{2}-t^{}_{1})}\ . 
 \label{con1}
 \end{align}
As a result, 
 \begin{align}
\sum_{{\bf k}}G^{<}_{{\bf k}{\bf k}}(t,t)
= i\sum_{\bf k}f^{}_{L}(\epsilon^{}_{k})&\Big (1-\int ^{t}dt^{}_{2}\int^{t^{}_{2}} dt^{}_{1}e^{-i\epsilon^{}_{k}(t^{}_{2}-t^{}_{1})}
 \oJ^{}_{L}(t^{}_{1})G^{a}_{dd}(t^{}_{1},t^{}_{2})\oJ^{\dagger}_{L}(t^{}_{2})\nonumber\\
 &+\int ^{t}dt^{}_{1}\int^{t^{}_{1}} dt^{}_{2}e^{-i\epsilon^{}_{k}(t^{}_{2}-t^{}_{1})}
 \oJ^{}_{L}(t^{}_{1})G^{r}_{dd}(t^{}_{1},t^{}_{2})\oJ^{\dagger}_{L}(t^{}_{2})\Big )\nonumber\\
 &+ \sum_{\bf k}\int ^{t}dt^{}_{1}\int^{t}dt^{}_{2}e^{-i\epsilon^{}_{k}(t^{}_{2}-t^{}_{1})}
 \oJ^{}_{L}(
  t^{}_{1})G^{<}_{dd}(t^{}_{1},t^{}_{2})\oJ^{\dagger}_{L}(t^{}_{2})\ .
  \label{con2}
\end{align} 
The first two integrals result in terms linear in $t$ (this would be the sole result when the time dependence of the tunneling amplitudes is ignored), and terms behaving as $\exp[\pm i\Omega t]/(\pm i\Omega)$. The third integrand includes [see Eq.~(\ref{Gdo})]
$\sum_{{\bf k}'}f^{}_{L}(\epsilon^{}_{{k}'})\exp[i(\epsilon^{}_{k}-\epsilon^{}_{k'})(t^{}_{1}-t^{}_{2})]$, where ${\bf k}$ can be equal to ${\bf k}'$, in which case a quadratic dependence on $t$ or  $\exp[\pm i\Omega t]/(\pm i\Omega)$ will appear, or ${\bf k}\neq{\bf k}'$. To avoid these complications, it is customary to consider the time derivative of the equal-momentum Green's function of the reservoirs.

%%%%%%%%%%%%%%%%%%%%%%%%%%%%%%%%%%%%%%%%%%%%%%%%%%%%%%%%%%%
%%%%%%%%%%%%%%%%%%%%%%%%%%%%%%%%%%%%%%%%%%%%%%%%%%%%%%%%%%%

%\subsection {The spin densities}

\subsection {The magnetization}

\label{leadspin}

%%%%%%%%%%%%%%%%%%%%%%%%%%%%%%%%%%%%%%%%%%%%%%%%%%%%%%%%%%%
%%%%%%%%%%%%%%%%%%%%%%%%%%%%%%%%%%%%%%%%%%%%%%%%%%%%%%%%%%%

Here, we summarize our results for the magnetization of the various parts of the device considered.
The magnetization generated on the quantum dot, exploiting Eqs. (\ref{dotM}) and (\ref{Gdtt}), is
\begin{align}
{\bf M}_{d}(t)&=
 \frac{2\Omega\Gamma^{}_{L}}{\pi}\int d\epsilon^{}_{k}f^{}_{L}(\epsilon^{}_{k})\frac{2(\epsilon^{}_{d}-\epsilon^{}_{k})S^{}_{L}}{[(\epsilon^{}_{d}-\epsilon^{}_{k})^{2}+\Gamma^{2}]^{2}}[S^{}_{L}\hat{\bf x}-C^{}_{L}\hat{\bf n}(t)]\nonumber\\
 &+\frac{2\Omega\Gamma^{}_{R}}{\pi}\int d\epsilon^{}_{p}f^{}_{R}(\epsilon^{}_{p})\frac{2(\epsilon^{}_{d}-\epsilon^{}_{p})S^{}_{R}}{[(\epsilon^{}_{d}-\epsilon^{}_{p})^{2}+\Gamma^{2}]^{2}}[S^{}_{R}\hat{\bf x}+C^{}_{R}\hat{\bf n}(t)]\ \ .
\end{align}
The magnetization  created in the leads by the momentum non-conserving processes is related to the 
magnetization flux 
due to momentum-conserving processes,
\begin{align}
\dot{\bf M}^{\rm con}_{L(R)}(t)={\bf M}^{\rm non-con}_{L(R)}(t)/(\pi{\cal N}^{}_{L(R)})\ ,
\end{align}
where 
\begin{align}
\dot{\bf M}^{\rm con}_{L}(t)&\approx\frac{2\Omega\Gamma^{}_{L}\Gamma^{}_{R}}{\pi}\Big(\int d\epsilon^{}_{k}f^{}_{L}(\epsilon^{}_{k})\frac{2(\epsilon^{}_{d}-\epsilon^{}_{k})S^{}_{L}}{[(\epsilon^{}_{d}-\epsilon^{}_{k})^{2}+\Gamma^{2}]^{2}}{\rm Tr}\{\sig[S^{}_{L}\hat{\bf x}+C^{}_{L}\hat{\bf n}(t)]\cdot\sig\}
\nonumber\\
&+\int d\epsilon^{}_{p}f^{}_{R}(\epsilon^{}_{p})\frac{2(\epsilon^{}_{d}-\epsilon^{}_{p})S^{}_{R}}{[(\epsilon^{}_{d}-\epsilon^{}_{p})^{2}+\Gamma^{2}]^{2}}{\rm Tr}\{\sig[S^{}_{2L+R}\hat{\bf x}+C^{}_{2L+R}\hat{\bf n}(t)]\cdot\sig\}\Big )\ ,
\end{align}
and
\begin{align}
\dot{\bf M}^{\rm con}_{R}(t)&\approx\frac{2\Omega\Gamma^{}_{L}\Gamma^{}_{R}}{\pi}\Big (\int d\epsilon^{}_{p}f^{}_{R}(\epsilon^{}_{p})\frac{2(\epsilon^{}_{d}-\epsilon^{}_{p})S^{}_{R}}{[(\epsilon^{}_{d}-\epsilon^{}_{p})^{2}+\Gamma^{2}]^{2}}{\rm Tr}\{\sig[S^{}_{R}\hat{\bf x}-C^{}_{R}\hat{\bf n}(t)]\cdot\sig\}\nonumber\\
&+\int d\epsilon^{}_{k}f^{}_{L}(\epsilon^{}_{k})\frac{2(\epsilon^{}_{d}-\epsilon^{}_{k})S^{}_{L}}{[(\epsilon^{}_{d}-\epsilon^{}_{k})^{2}+\Gamma^{2}]^{2}}{\rm Tr}\{\sig[S^{}_{2R+L}\hat{\bf x}-C^{}_{2R+L}\hat{\bf n}(t)]\cdot\sig\}\Big )\ .
\end{align}
In particular, 
\begin{align}
&\dot{\bf M}^{\rm con}_{L}(t)+\dot{\bf M}^{\rm con}_{R}(t)\approx\frac{4\Omega\Gamma^{}_{L}\Gamma^{}_{R}\sin(k^{}_{\rm so}[d^{}_{L}+d^{}_{R}])}{\pi}\nonumber\\
&\times
\Big(\int d\epsilon^{}_{k}f^{}_{L}(\epsilon^{}_{k})\frac{2(\epsilon^{}_{d}-\epsilon^{}_{k})\sin(k^{}_{\rm s}d^{}_{L})}{[(\epsilon^{}_{d}-\epsilon^{}_{k})^{2}+\Gamma^{2}]^{2}}{\rm Tr}\{\sig[\cos(k^{}_{\rm so}d^{}_{R})\hat{\bf x}+\sin(k^{}_{\rm so}d^{}_{R})\hat{\bf n}(t)]\cdot\sig\}\nonumber\\
&+\int d\epsilon^{}_{p}f^{}_{R}(\epsilon^{}_{p})\frac{2(\epsilon^{}_{d}-\epsilon^{}_{p})\sin(k^{}_{\rm so}d^{}_{R})}{[(\epsilon^{}_{d}-\epsilon^{}_{p})^{2}+\Gamma^{2}]^{2}}{\rm Tr}\{\sig[\cos(k^{}_{\rm so}d^{}_{L})\hat{\bf x}-\sin(k^{}_{\rm so}d^{}_{L})\hat{\bf n}(t)]\cdot\sig\}\Big )\ .
\end{align}

%%%%%%%%%%%%%%%%%%%%%%%%%%%%%%%%%%%%%%%%%%%%%%%%%%%%%%%%%%%
%%%%%%%%%%%%%%%%%%%%%%%%%%%%%%%%%%%%%%%%%%%%%%%%%%%%%%%%%%%

%%%%%%%%%%%%%%%%%%%%%%%%%%%%%%%%%%%%%%%%%%%%%%%%%%%%%%%%%%%
%%%%%%%%%%%%%%%%%%%%%%%%%%%%%%%%%%%%%%%%%%%%%%%%%%%%%%%%%%%

\subsection{ The particle current}
\label{particlef}

%%%%%%%%%%%%%%%%%%%%%%%%%%%%%%%%%%%%%%%%%%%%%%%%%%%%%%%%%%%
%%%%%%%%%%%%%%%%%%%%%%%%%%%%%%%%%%%%%%%%%%%%%%%%%%%%%%%%%%%

The particle current is given by tracing Eq.~(\ref{Gkkdt}). The trace of the first term is
\begin{align}
%-i\sum_{{\bf k}}f^{}_{L}(\epsilon^{}_{k}){\rm Tr}\{\oJ^{}_{L,+}(\epsilon^{}_{k},t)\oJ^{\dagger}_{L}(t)&+\oJ^{}_{L}(t)\oJ^{}_{L,-}(\epsilon^{}_{k},t)\}%={\rm Tr}\{-i|\oJ^{}_{L,0}|^{2}\Big (C^{}_{L}+iS^{}_{L}[\sigma^{}_{-}e^{i\Omega t}+\sigma^{}_{+}e^{-i\Omega t}\Big)\nonumber\\
%&\times\Big (\frac{C^{}_{L}}{\epsilon^{}_{d}-\epsilon^{}_{k}-i\Gamma}%\nonumber\\
%-iS^{}_{L}\Big [\frac{\sigma^{}_{-}e^{i\Omega t}}{\epsilon^{}_{d}-\epsilon^{}_{k}+\Omega-i\Gamma}+\frac{\sigma^{}_{+}e^{-i\Omega t}}{\epsilon^{}_{d}-\epsilon^{}_{k}-\Omega-i\Gamma}\Big]\Big )+{\rm H.c.}\}=\nonumber\\
-4i\Gamma|J^{}_{L,0}|^{2}\sum_{\bf k}f^{}_{L}(\epsilon^{}_{k})\Big (\frac{C^{2}_{L}}{(\epsilon^{}_{d}-\epsilon^{}_{k})^{2}+\Gamma^{2}}%\nonumber\\
%&+\frac{S^{2}_{L}}{2}\Big [
%\frac{1}{(\epsilon^{}_{d}-\epsilon^{}_{k}-\Omega)^{2}+\Gamma^{2}}+\frac{1}{(\epsilon^{}_{d}-\epsilon^{}_{k}+\Omega)^{2}+\Gamma^{2}}\Big]\Big)\nonumber\\&
+S^{2}_{L}\frac{(\epsilon^{}_{d}-\epsilon^{}_{k})^{2}+\Omega^{2}+\Gamma^{2}}{[
(\epsilon^{}_{d}-\epsilon^{}_{k}-\Omega)^{2}+\Gamma^{2}][
(\epsilon^{}_{d}-\epsilon^{}_{k}+\Omega)^{2}+\Gamma^{2}]}\Big )\ , 
\end{align}
and that of the second term is [see Eq. (\ref{Gdo})]
\begin{align}
%&2\pi{\cal N}^{}_{L}|\oJ^{}_{L,0}|^{2}{\rm Tr}\{G^{<}_{dd}(t,t)\}%=2i\Gamma^{}_{L}\Big(\sum_{\bf k}f^{}_{L}(\epsilon^{}_{k}){\rm Tr}\{\oJ^{}_{L,-}(\epsilon^{}_{k},t)\oJ^{\dagger}_{L,+}(\epsilon^{}_{k},t)\}+\sum_{\bf p}f^{}_{R}(\epsilon^{}_{p}){\rm Tr}\{\oJ^{}_{R,-}(\epsilon^{}_{p},t)\oJ^{\dagger}_{R,+}(\epsilon^{}_{p},t)\}\Big)\nonumber\\
%&=2i\Gamma^{}_{L}|\oJ^{}_{L,0}|^{2}{\rm Tr}\Big\{\Big (\frac{C^{}_{L}}{\epsilon^{}_{d}-\epsilon^{}_{k}-i\Gamma}
%-iS^{}_{L}\Big [\frac{\sigma^{}_{-}e^{i\Omega t}}{\epsilon^{}_{d}-\epsilon^{}_{k}+\Omega-i\Gamma}+\frac{\sigma^{}_{+}e^{-i\Omega t}}{\epsilon^{}_{d}-\epsilon^{}_{k}-\Omega-i\Gamma}\Big]\Big )\nonumber\\
%&\times\Big (\frac{C^{}_{L}}{\epsilon^{}_{d}-\epsilon^{}_{k}+i\Gamma}
%+iS^{}_{L}\Big [\frac{\sigma^{}_{-}e^{i\Omega t}}{\epsilon^{}_{d}-\epsilon^{}_{k}+\Omega+i\Gamma}+\frac{\sigma^{}_{+}e^{-i\Omega t}}{\epsilon^{}_{d}-\epsilon^{}_{k}-\Omega+i\Gamma}\Big]\Big )\Big\}\nonumber\\=
&4i\Gamma^{}_{L}|J^{}_{L,0}|^{2}\sum_{\bf k}f^{}_{L}(\epsilon^{}_{k})\Big (\frac{C^{2}_{L}}{(\epsilon^{}_{d}-\epsilon^{}_{k})^{2}+\Gamma^{2}}+S^{2}_{L}\frac{(\epsilon^{}_{d}-\epsilon^{}_{k})^{2}+\Omega^{2}+\Gamma^{2}}{[
(\epsilon^{}_{d}-\epsilon^{}_{k}-\Omega)^{2}+\Gamma^{2}][
(\epsilon^{}_{d}-\epsilon^{}_{k}+\Omega)^{2}+\Gamma^{2}]}\Big ) \nonumber\\
&+4i\Gamma^{}_{L}|J^{}_{R,0}|^{2}\sum_{\bf p}f^{}_{R}(\epsilon^{}_{p})\Big (\frac{C^{2}_{R}}{(\epsilon^{}_{d}-\epsilon^{}_{p})^{2}+\Gamma^{2}}+S^{2}_{R}\frac{(\epsilon^{}_{d}-\epsilon^{}_{p})^{2}+\Omega^{2}+\Gamma^{2}}{[
(\epsilon^{}_{d}-\epsilon^{}_{p}-\Omega)^{2}+\Gamma^{2}][
(\epsilon^{}_{d}-\epsilon^{}_{p}+\Omega)^{2}+\Gamma^{2}]}\Big )\ .
\end{align}
Noting that $\Gamma=\Gamma^{}_{L}+\Gamma^{}_{R}$, 
the particle flux from the left lead due to the momentum-conserving processes becomes
steady. i.e., does not depend on time. To the lowest possible order in $\Omega$, it reads 
\begin{align}
&I^{}_{L}=-\frac{4\Gamma^{}_{L}\Gamma^{}_{R}}{\pi}\Big (\int d\epsilon^{}_{k}\frac{f^{}_{L}(\epsilon^{}_{k})}{(\epsilon^{}_{d}-\epsilon^{}_{k})^{2}+\Gamma^{2}}-\int d\epsilon^{}_{p}\frac{f^{}_{R}(\epsilon^{}_{p})}{(\epsilon^{}_{d}-\epsilon^{}_{p})^{2}+\Gamma^{2}}\Big )\nonumber\\
&%-\frac{8\Gamma^{2}_{L}\Omega^{2}S^{2}_{L}}{\pi}\int d\epsilon^{}_{k}\frac{
%f^{}_{L}(\epsilon^{}_{k})}{[
%(\epsilon^{}_{d}-\epsilon^{}_{k})^{2}+\Gamma^{2}]^{2}}
+\frac{4\Gamma^{}_{L}\Gamma^{}_{R}\Omega^{2}}{\pi}\Big (\int d\epsilon^{}_{k}\frac{S^{2}_{L}f^{}_{L}(\epsilon^{}_{k})}{[(\epsilon^{}_{d}-\epsilon^{}_{k})^{2}+\Gamma^{2}]^{2}}-\int d\epsilon^{}_{p}\frac{S^{2}_{R}f^{}_{R}(\epsilon^{}_{p})}{[(\epsilon^{}_{d}-\epsilon^{}_{p})^{2}+\Gamma^{2}]^{2}}\Big )\ .
\end{align}
This result was derived and discussed in Ref. \onlinecite{O2020}, and is included here for completeness.

%%%%%%%%%%%%%%%%%%%%%%%%%%%%%%%%%%%%%%%%%%%%%%%%%%%%%%%%%%%
%%%%%%%%%%%%%%%%%%%%%%%%%%%%%%%%%%%%%%%%%%%%%%%%%%%%%%%%%%%

 %%%%%%%%%%%%%%%%%%%%%%%%%%%%%%%%%%%%%%%%%%%%%%%%%%%%%%%%%%%

\twocolumngrid


\begin{thebibliography}{99}
\bibitem{SOIcomment}
 Alternatively, the Rashba coefficient $\alpha_R = \hbar^2 k_{\rm so}/m^*$ can be used to characterize the strength of the spin-orbit interaction.
\bibitem{vFcomment}
 The Fermi velocity is high for a semiconductor, presumably because the electric field created by the side gates 
induce a high electron density in the InAs nanowire.
\end{thebibliography}
\end{document}